\documentclass[twocolumn]{aastex62}
\submitjournal{ApJ}
\usepackage{enumitem}
\usepackage{graphicx,url}
\usepackage{color}

\def\mrk509{Mrk~509}
\def\MYtorus{{\tt MYtorus}}
\def\Borus{{\tt borus02}}
\def\nthComp{{\tt nthComp}}

\def\xmm{{\it XMM-Newton}}
\def\suzaku{{\it Suzaku}}
\def\swift{{\it Swift}}
\def\nustar{{\it NuSTAR}}
\def\integral{{\it INTEGRAL}}
\def\chandra{{\it Chandra}}
\def\xillver{{\tt xillver}}
\def\xillverD{{\tt xillverD}}
\def\xillverCp{{\tt xillverCp}}

\def\relxillD{{\tt relxillD}}

\def\relxilllpD{{\tt relxilllpD}}

\def\msun{M_{\odot}}

\shorttitle{The Soft-Excess in Mrk 509: Warm Corona or Relativistic Reflection?}
\shortauthors{Garc\'{\i}a \& et al.}

\begin{document}

\title{\large\bf The Soft-Excess in Mrk 509: Warm Corona or Relativistic Reflection?}




\correspondingauthor{Javier~A.~Garc\'ia}
\email{javier@caltech.edu}

\author[0000-0003-3828-2448]{Javier~A.~Garc\'ia}
\affil{Cahill Center for Astronomy and Astrophysics, California Institute of Technology, Pasadena, CA 91125, USA}
\affil{Dr. Karl Remeis-Observatory and Erlangen Centre for Astroparticle Physics, Sternwartstr.~7, 96049 Bamberg, Germany}

\author{Erin~Kara}
\affil{Department of Astronomy, University of Maryland, College Park, MD 20742, USA}

\author{Dominic~Walton}
\affil{Institute of Astronomy, Madingley Road, Cambridge CB3 0HA, UK}

\author{Tobias~Beuchert}
\affil{Anton Pannekoek Institute for Astronomy, Universiteit van Amsterdam, Science Park 904, 1098 XH, Amsterdam, The Netherlands}

\author{Thomas~Dauser}
\affil{Dr. Karl Remeis-Observatory and Erlangen Centre for Astroparticle Physics, Sternwartstr.~7, 96049 Bamberg, Germany}

\author{Efrain~Gatuzz}
\affil{ESO, Karl-Schwarzschild-Strasse 2, D-85748 Garching bei M\"unchen, Germany}
\affil{Excellence Cluster Universe, Boltzmannstr. 2, D-85748, Garching, Germany}

\author{Mislav~Balokovic}
\affil{Harvard-Smithsonian Center for Astrophysics, 60 Garden Street, Cambridge, MA 02140, USA}

\author{James~F.~Steiner}
\affil{MIT Kavli Institute for Astrophysics and Space  Research, MIT, 70 Vassar Street, Cambridge, MA 02139}

\author{Francesco~Tombesi}
\affil{Dipartimento di Fisica, Universit\`a di Roma ``Tor Vergata", via della Ricerca Scientifica 1, I-00133, Roma, Italy}
\affil{INAF Astronomical Observatory of Rome, Via Frascati 33, 00078 Monteporzio Catone, Italy}
\affil{NASA/Goddard Space Flight Center, Code 662, Greenbelt, MD 20771, USA}
\affil{Department of Astronomy, University of Maryland, College Park, MD 20742, USA}

\author{Riley~M.~T.~Connors}
\affil{Cahill Center for Astronomy and Astrophysics, California Institute of Technology, Pasadena, CA 91125, USA}

\author{Timothy~R.~Kallman}
\affil{NASA/Goddard Space Flight Center, Code 662, Greenbelt, MD 20771, USA}

\author{Fiona~A.~Harrison}
\affil{Cahill Center for Astronomy and Astrophysics, California Institute of Technology, Pasadena, CA 91125, USA}

\author{Andrew~Fabian}
\affil{Institute of Astronomy, Madingley Road, Cambridge CB3 0HA, UK}

\author{J\"orn~Wilms}
\affil{Dr. Karl Remeis-Observatory and Erlangen Centre for Astroparticle Physics, Sternwartstr.~7, 96049 Bamberg, Germany}

\author{Daniel~Stern}
\affil{Jet Propulsion Laboratory, California Institute of Technology, 4800 Oak Grove Drive, Mail Stop 169-221, Pasadena, CA 91109, USA}

\author{Lauranne~Lanz}
\affil{Department of Physics and Astronomy, Dartmouth College, 6127 Wilder Laboratory, Hanover, NH 03755, USA}

\author{Claudio~Ricci}
\affil{N\'ucleo de Astronom\'ia de la Facultad de Ingenier\'ia, Universidad Diego Portales, Av. Ej\'ercito Libertador 441, Santiago, Chile}
\affil{Kavli Institute for Astronomy and Astrophysics, Peking University, Beijing 100871, China}
\affil{Chinese Academy of Sciences South America Center for Astronomy, Camino El Observatorio 1515, Las Condes, Santiago, Chile}

\author{David~R.~Ballantyne}
\affil{Center for Relativistic Astrophysics, School of Physics, Georgia
  Institute of Technology, 837 State Street, Atlanta, GA 30332-0430}

\begin{abstract}
%
We present the analysis of the first \nustar\ observations ($\sim 220$\,ks),
simultaneous with the last \suzaku\ observations ($\sim 50$\,ks), of the active
galactic nucleus of the bright Seyfert 1 galaxy \mrk509. 
The time-averaged spectrum in the $1-79$\,keV X-ray
band is dominated by a power-law continuum ($\Gamma\sim 1.8-1.9$), a strong
soft excess around 1\,keV, and signatures of X-ray reflection in the form of
Fe K emission ($\sim 6.4$\,keV), an Fe K absorption edge ($\sim
7.1$\,keV), and a Compton hump due to electron scattering ($\sim 20-30$\,keV).
We show that these data can be described by two very different prescriptions for
the soft excess: a warm ($kT\sim 0.5-1$\,keV) and optically thick ($\tau\sim10-20$)
Comptonizing corona, or a relativistically blurred ionized reflection spectrum
from the inner regions of the accretion disk. While these two scenarios cannot
be distinguished based on their fit statistics, we argue that the parameters 
required by the warm corona model are physically incompatible with the conditions
of standard coronae. Detailed photoionization calculations show that
even in the most favorable conditions, the warm corona should produce strong
absorption in the observed spectrum. On the other hand, while the relativistic
reflection model provides a satisfactory description of the data, it also requires
extreme parameters, such as maximum black hole spin, a very low and compact hot
corona, and a very high density for the inner accretion disk. Deeper observations
of this source are thus necessary to confirm the presence of relativistic 
reflection, and to further understand the nature of its soft excess.
\end{abstract}


%
%
%
%
\section{Introduction}\label{sec:intro}

Accretion onto supermassive black holes in active galactic nuclei (AGN) is one
of the most efficient mechanisms to convert gravitational energy into
radiation, comprised mostly of very energetic photons. For this
reason, X-ray spectroscopy is a resourceful  technique to study supermassive
black holes and their interaction with their surroundings. In the case of most
Seyfert AGN, the X-ray continuum is typically dominated by a power-law that
extends to high energies, which is thought to be produced either in a
central hot corona \citep[e.g., ][]{sha73,haa93}, or at the base of a jet
\citep[e.g.,][]{mat92,mar05}, although the exact mechanism is still a matter of
study.  Thermal emission from the accretion disk peaks in the ultraviolet (UV)
band, extending partially to the soft X-rays.  A fraction of the coronal
emission illuminates the accretion disk, producing a rich reflection spectrum
of fluorescent lines and other spectral features. This reflection component
can be ionized, as changes in the ionization state of the disk determine spectral
features observed \citep[e.g.,][]{ros05,gar10}, and also blurred and distorted by relativistic effects
\citep[e.g.,][]{lao91,cru06}, if it originates close enough
to the supermassive black hole; or it can be cold and neutral, if produced farther from the 
black hole in either the broad-line region or the torus \citep[e.g.,][]{mat91,geo91}. 

In a large fraction of Seyfert AGN a soft excess component is also observed
peaking near 1--2\,keV. Its origin has been debated over the years. This
soft excess was first believed to be the hard tail of UV blackbody emission
from the accretion disk \citep{sin85,arn85,pou86,mag98,lei99}; however, this
explanation was ruled out given that systems with very different accretion rates
and/or masses would be characterized by the same blackbody temperature, which is
not expected for an accretion disk \citep{gie04b,por04,pic05,min09}.  The
current models invoked to explain the soft excess tend to favor either 
Comptonization of UV photons or blurred ionized reflection. In the first case, the
disk photons are Comptonized by a corona above the disk, which is optically
thicker and cooler than the corona responsible for the primary X-ray emission
\citep{cze87,mid09,jin09,don12}. In the second case, the emission lines produced 
in the disk are relativistically blurred due to the proximity to the
black hole \citep{fab02,ros05,cru06,gar10,wal13}.  

The Seyfert type 1 galaxy Mrk~509 was one of the first AGN to be
studied in detail because it is luminous
\citep[$L_\mathrm{Bol}=1.07\times 10^{45}$\,erg\,s$^{-1}$;][]{woo02}
and relatively nearby \citep[$z=0.0344$;][]{fis95}.  The corresponding
X-ray flux of $F_\mathrm{x} = (2$--$5)\times
10^{-11}$\,erg\,cm$^{-2}$\,s$^{-1}$ \citep{kaa11} is powered by a
$1.4\times 10^8\,M_{\odot}$ black hole \citep{pet04}, which is
accreting at 20--30\% of the Eddington rate \citep{pet13}. Excess soft
($\lesssim2$\,keV) emission above the extrapolation of the hard X-ray
continuum was first identified by \cite{sin85}. After \citet{mor87}
detected the Fe line, improved X-ray instruments and detectors led to
a full discussion of reflection features by \citet{pou94}.

An intense campaign of multi-wavelength monitoring of Mrk~509 involving the
X-ray observatories {\it XMM-Newton} and {\it Suzaku} has provided a detailed
model for the observed set of soft-X-ray absorption features, caused by
differentially ionized warm absorbing gas \citep{kaa11}. Portions of this gas
phase have been observed to be outflowing at different velocities
\citep{smi07}, including a component classed as an ultra-high velocity outflow
\citep{cap09}. This campaign also resulted in the most complete study of the Fe
K complex of Mrk~509 to date, revealing a neutral narrow component and an
ionized broad component. The latter has been interpreted as relativistic
reflection from the inner regions of the accretion disk \citep{wal13}. Despite
the presence of a warm absorber, Mrk~509 can still be considered a ``bare" AGN.
The intrinsic absorption is low enough that it does not complicate
determination of the reflection continuum \citep{wal13}.

Most of the previous analyses of Mrk~509 mentioned above have predominantly
focused on understanding the physical details of the warm absorber,
the soft excess and the high-velocity outflows. Our emphasis
is upon the detection or not of {\sl relativistic reflection}
features, namely the Fe K complex and the Compton hump, which are likely
to originate due to the reprocessing of hard X-rays in the inner-most
regions of the accretion disk.  To date,
observations of the hard X-ray component in which these signatures are
most evident are quite limited, and the physical picture is
accordingly subject to large and fundamental uncertainties
\citep[e.g.,][]{pet13,pon13,kaa14}.

%
\begin{deluxetable*}{cccccc}[ht!]
\tablecaption{Observational Data Log for \mrk509 \label{tab:obslog}}
\tablecolumns{6}
\tablewidth{0pt}
\tablehead{
\colhead{Telescope} & \colhead{Instrument} & \colhead{ObsId} & \colhead{Date} &
\colhead{Exp (ks)} & \colhead{Counts ($10^5$)} }
\startdata
\nustar\  & FPMA/B   & 60101043002 & 2015-04-29 & 166  & 3.2  \\
\nustar\  & FPMA/B   & 60101043004 & 2015-06-02 &  37  & 0.6  \\
\suzaku\  & XIS~0/3  & 410017010   & 2015-05-01 &  47  & 2.1  \\
\enddata
\end{deluxetable*}
%

The {\it Nuclear Spectroscopic Telescope Array} \citep[\nustar,][]{har13} low
background, high sensitivity and $\sim3$--$79$\,keV bandwidth (which captures
the key reflection features), together with the development of advanced
relativistic reflection models such as {\tt relxill}
\citep{dau13,gar13a,gar14a}, have revolutionized studies of X-ray reflection
spectroscopy \citep[e.g.,][]{ris13,wal14,kec14,kar17,por18}.  In this paper, we
present analysis of the first \nustar\ and the last \suzaku\ observations of
the bright AGN \mrk509. Implementing a variety of X-ray spectral models, we
investigate the origin of the soft excess and the possibility for relativistic
ionized reflection in this source.  Based on these fits, we present a
theoretical discussion on the physical implications of two competing models to
explain the soft excess: the warm corona and the relativistic reflection.

\section{Observational Data}\label{sec:obs}

The first \nustar\ observations of \mrk509\ were taken during Cycle~1
of the Guest Observer Program on 2015-April-29, with a total requested exposure
time of 200\,ks. A simultaneous \suzaku\ observation was performed with
a 50\,ks exposure in order to provide low energy coverage. The
\nustar\ exposure was interrupted after $\sim$165\,ks due to a Target
of Opportunity trigger. The remaining $\sim$35\,ks were taken roughly a
month later on 2015-June-02. A log with details of the
observational data analyzed in this paper is shown in
Table~\ref{tab:obslog}.

\subsection{NuSTAR Extraction}

The \nustar\ data are split over two ObsId, 60101043002 and
60101043004, separated by roughly a month. We reduced these data following
standard procedures using the \nustar\ Data Analysis Software
({\small NUSTARDAS}, v1.6.0) and instrumental calibration files from caldb
v20160824. We first cleaned the unfiltered event files with {\small
NUPIPELINE}. We used the standard depth correction, which significantly
reduces the internal high-energy background, and also removed passages
through the South Atlantic Anomaly, again using standard filtering
parameters. Source and background spectra/lightcurves and instrumental
responses were then produced for both focal plane modules FPMA and FPMB
using {\small NUPRODUCTS}. Source products were extracted from circular
regions of radius 120$''$, and background was estimated from regions of
blank sky on the same detector as \mrk509. In order to maximize the
signal-to-noise (S/N), in addition to the standard `science' (mode 1) data,
we also extracted the `spacecraft science' (mode 6) data following
\cite{wal16}. In this case, the mode 6 data provide $\sim$10\% of the
total $\sim$220\,ks good \nustar\ exposure. 

%
\graphicspath{{/Users/javier/mrk509/march-2017/data_for_javier/lightcurves/}}
\begin{figure*}[ht!]
\centering
\includegraphics[width=\linewidth]{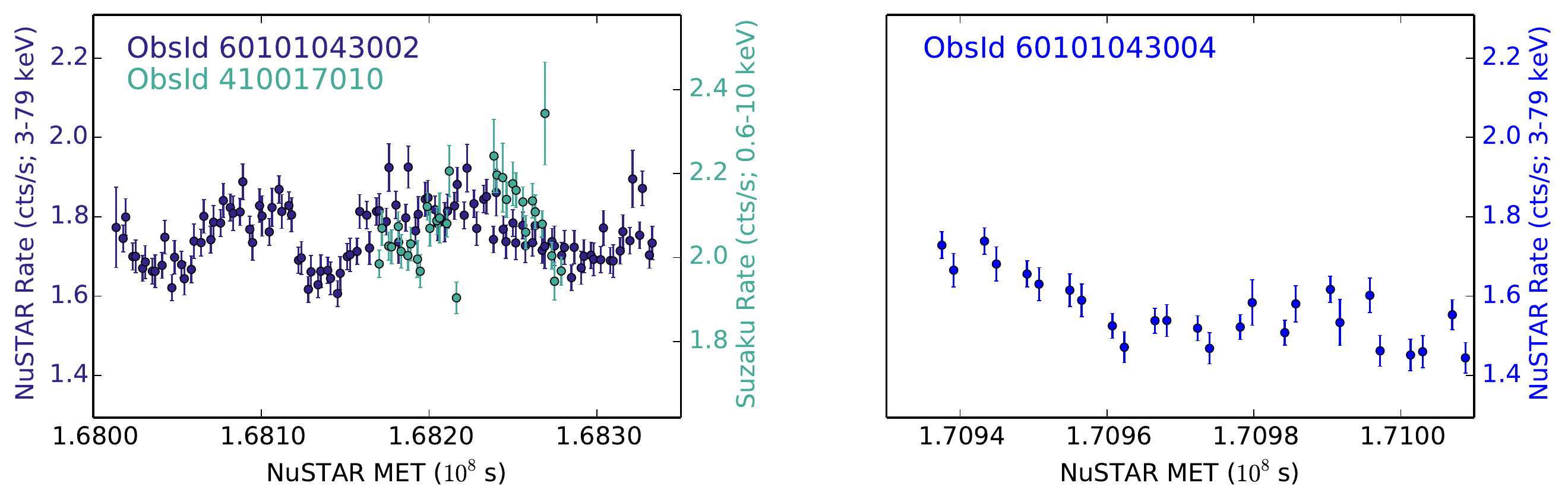}
\caption{Light curves for the \nustar\ FPMA and \suzaku\ XIS exposures of
\mrk509\ binned in 2\,ks intervals. The left panel shows the first
$\sim$165\,ks \nustar\ exposure (ObsId 60101043002), together with the
$\sim$50\,ks \suzaku\ exposure (ObsId 410017010), while the right panel shows
the shorter $\sim$65\,ks \nustar\ observation (ObsId 60101043004).  The source
shows a very stable count rate through the full range, with no obvious flares
or dips.
}
\label{fig:lightcurves}
\end{figure*}

\subsection{Suzaku Extraction}

The \suzaku\ data were reduced starting from the unfiltered event files and 
then screened applying the standard selection criteria described in the
\suzaku\ ABC
guide\footnote{\url{http://heasarc.gsfc.nasa.gov/docs/suzaku/analysis/abc/}}.
The source spectra were extracted from circular regions of 2.5$\arcmin$ radius
centered on the source, whereas background spectra were extracted from a region
of the same size offset from the main target and avoiding the calibration
sources.  We generated the redistribution matrix file (RMF) and the ancillary 
response file (ARF) of the X-ray Imaging Spectrometer (XIS) with the
\texttt{xisrmfgen} and \texttt{xissimarfgen} ftools, respectively. We selected
the XIS data in both the 3$\times$3 and 5$\times$5 modes. The spectra were
inspected for possible pile-up contamination and this possibility was excluded.
The spectra of the front-illuminated XIS instruments (XIS~0 and XIS~3) were
merged after checking that their fluxes were consistent. The data from the
back-illuminated XIS instrument, XIS~1, is not used due to its much lower
sensitivity in the Fe K band and cross-calibration uncertainties with the
front-illuminated XIS~0 and XIS~3.

\subsection{Light Curves and Time-Averaged Spectra}\label{sec:data}

The lightcurves for the two \nustar\ and the \suzaku\ exposures are 
shown in Figure~\ref{fig:lightcurves}. The data were binned in 2\,ks
intervals. The \suzaku\ exposure is simultaneous with the first and longer
\nustar\ exposure. The lightcurves show a very similar level of variability,
which in both cases is very weak ($\sim 6$\%). This value corresponds to the
normalized excess variance \citep{vau03} that suppresses a possible
rms-flux correlation usually found for un-normalized rms measures.
The right panel of Figure~\ref{fig:lightcurves} contains the
lightcurve for the shorter \nustar\ exposure taken roughly a month
later. It shows a similar count rate with no significant variability;
neither flares nor strong dips are detected. Spectra extracted
from the two \nustar\ exposures imply consistency after visual
inspection. We therefore combined these into a single spectrum taking
advantage of the full $\sim$220\,ks exposure.

\begin{figure*}[ht!]
\centering
\includegraphics[width=\linewidth,trim={0.5cm 0 0.5cm 0}]{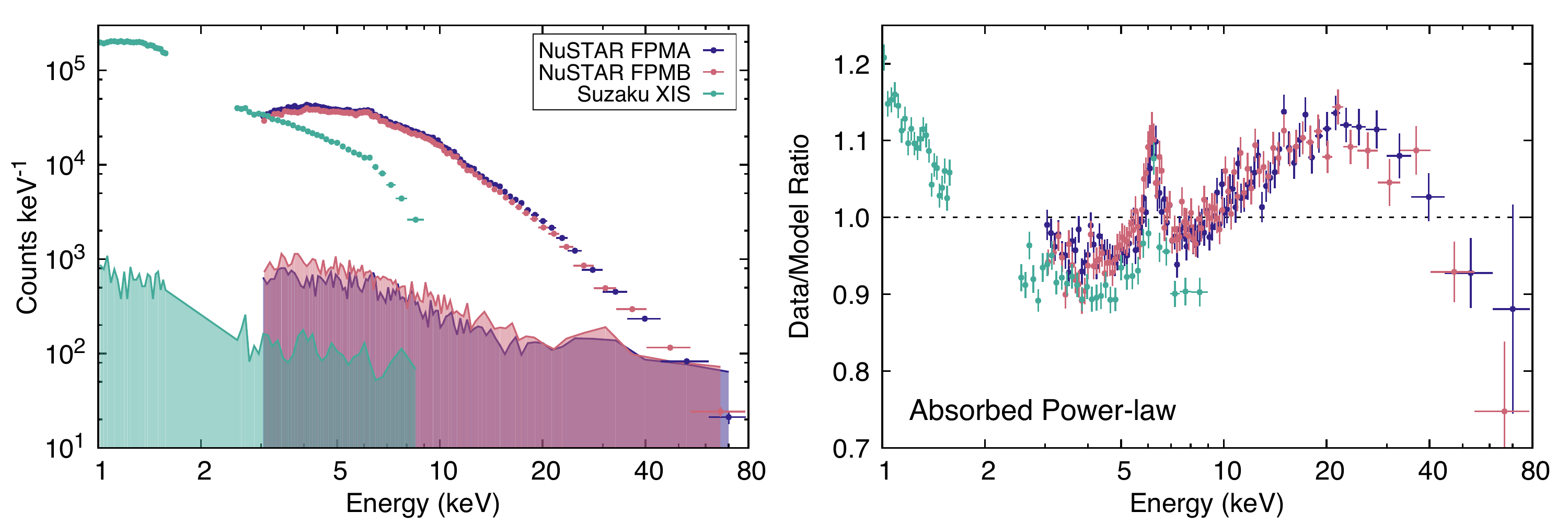}
\caption{Left: Count spectra from the 50\,ks \suzaku\ XIS (green) and the
220\,ks \nustar\ FPMA/B (blue/red) exposures. The shaded regions show the level
of background for each of the instruments. Right: Ratio of the data to an
absorbed power-law model, showing the soft excess at $\sim 1$\,keV, the Fe K
emission at $\sim 6.4$\,keV, and the Compton hump at $\sim 20-30$\,keV.
}
\label{fig:lcounts}
\end{figure*}

The final extracted total count spectra for \nustar's FPMA and FPMB, 
and \suzaku's XIS instruments are shown in
Figure~\ref{fig:lcounts}. The shaded region depicts the corresponding
backgrounds, which are well below the source counts up to $\sim$50\,keV. We
include \suzaku\ data in the 1--8\,keV range, excluding the 1.7--2.5\,keV range due
to calibration uncertainties.  We ignore data below 1\,keV due to concerns over
the quality of the calibration given molecular contamination of the XIS
detectors; contamination reduces the effective area differently on each detector, and as
a function of off-axis angle \citep{koy07,ket13}, and it is expected to worsen
over time \citep{mad17}.  \nustar\ data is included in the 3--79\,keV range.
The spectra were rebinned in order to oversample the instrument's resolution by
a factor of 3, and to ensure a minimum signal-to-noise of 6 per bin.

%
\graphicspath{{/Users/javier/mrk509/march-2017/plots/}}

\section{Spectral Analysis}\label{sec:anal}

We simultaneously fit the two \nustar\ FPMA and FPMB spectra extracted from the
full $\sim$220\,ks exposure together with the $\sim 50$\,ks \suzaku\ spectrum. The
fitting and statistical analysis presented here was carried out using the {\sc
xspec} package v-12.9.0d \citep{arn96}. A cross normalization constant is included 
to account for differences in the flux calibration among all three instruments (i.e.,
FPMA, FPMB, and XIS). The fitted values are consistent with those previously reported
by \cite{mad15}. All model parameter uncertainties are quoted to 90\%
confidence level.

Figure~\ref{fig:lcounts} (right)
shows the data-to-model ratio of these observations when fitted with a simple
absorbed power-law model (i.e., {\tt TBabs*pow}). The {\tt TBabs} component
is used to describe the Galactic absorption (see Section~\ref{sec:appro1}). The typical signatures of
X-ray reflection off optically thick material are evident in the spectrum: the 
fluorescent iron emission near 6.4\,keV, the iron K-edge near 7\,keV, and the 
Compton hump peaking at $\sim$25\,keV. Both instruments satisfactorily
agree in the shape and intensity of the iron emission. In absence of
relativistic effects, these features are well described by the reprocessing of
the X-rays in a relatively cold and neutral material, located far away from
the central region, possibly at the broad line region \citep[e.g.,][]{cos16,nar16}, or even
at the torus \citep[e.g.,][]{yaq07,mur09,mar18}.

The nature of the soft excess in \suzaku's bandpass, however, is not
yet very well established. As we shall show next, the particular
choice of the components used to model the soft excess has an
important effect in modeling of the reflected spectrum, and in
fact leads to different interpretations for this system. We will then
present fits with two different scenarios, and later discuss the
physical interpretation and implications for each one.

\subsection{Approach 1: Fitting the Soft Excess with a Warm Corona}\label{sec:appro1}

In their analysis of the \xmm\ and \integral\ campaign, \cite{pet13} proposed
that the clearly observed soft excess in \mrk509\ is due to the presence of a
warm corona, which they reproduced using a Comptonization model.
This corona can then be visualized as a warm ($kT_\mathrm{e}
\sim 0.5-1$\,keV) but optically thick ($\tau\sim 10-20$) atmosphere sitting on
top of the accretion disk.  This extended, slab-like corona is much colder than
the centrally located and possibly spherical corona responsible for the
power-law continuum emission that extends to high energies.  The emission of
the hot corona was fitted with a second Comptonization model, with a higher
coronal temperature ($kT_\mathrm{e} \sim 100$\,keV) and lower optical depth ($\tau\sim
0.5$).

We adopted the prescription of \cite{pet13} to fit the soft
excess. For this, we implemented two Comptonization components using
the \nthComp\ model \citep{zdz96,zyc99} with the required
parameters to reproduce the power-law continuum (hot corona) and
the soft excess (warm corona).  The hot corona component is
characterized by a slope of $\Gamma\sim 1.84$ and a relatively low
electron temperature of $kT_\mathrm{e}\sim30$\,keV. The temperature of the seed
photons for this component cannot be constrained and it is thus fixed to a 
relatively low value ($kT_\mathrm{BB}=100$\,eV).  The
warm corona component is characterized by a very soft continuum fixed
at $\Gamma=2.5$ and a much colder electron temperature $kT_\mathrm{e}\sim
0.4-0.5$\,keV, as well as a much lower temperature for the seed
photons, fixed at $kT_\mathrm{BB}=3$\,eV. The parameters held fixed
in these two components cannot be constrained with the current
dataset, likely due to the lack of data below 1\,keV. Their values
were chosen following the best-fit results of \cite{pet13}. The
intrinsic galactic absorption in this system is modeled using the {\tt
  TBabs} model with the corresponding abundances as set by
\cite{wil00}. This model automatically implements the \cite{ver96}
photoelectric cross sections. We freeze the column density to
$N_\mathrm{H}=4.25\times10^{20}$\,cm$^{-2}$ \citep{kal05}, and the source redshift to
$z=0.035$.

\begin{figure}[ht!]
\centering
\includegraphics[width=\linewidth,trim={0.5cm 1.5cm 0.5cm 0.5cm}]{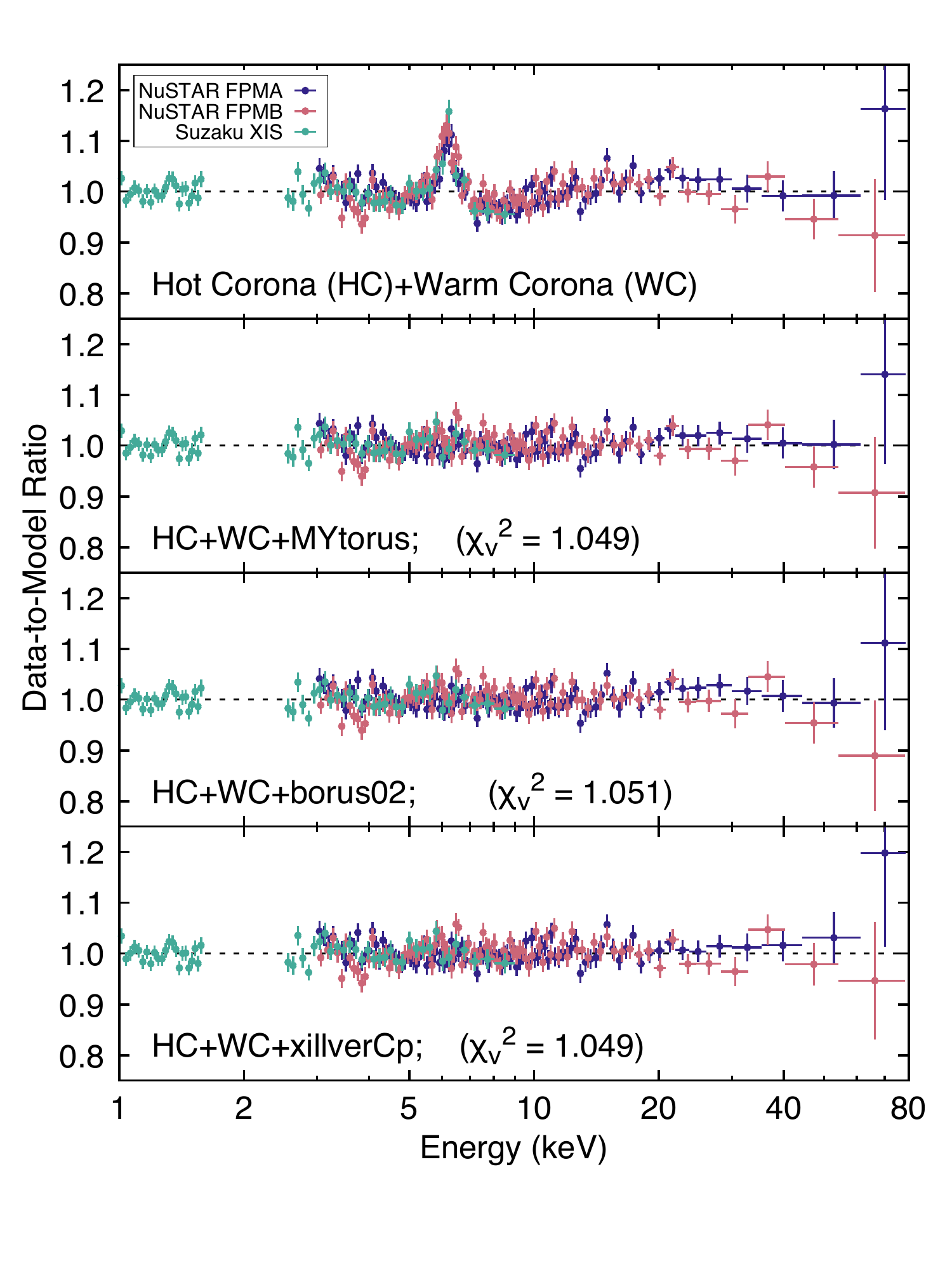}
\caption{Data-to-model ratios for the fits with the warm corona and different models
for the distant reflection.}
\label{fig:appro1}
\end{figure}

A data-to-model ratio plot of the fit using these models for the
continuum is shown in the top panel of Figure~\ref{fig:appro1}.  These two
Comptonization components, which are independent from one another, provide a
good fit to both the continuum and the soft excess, and the only obvious
residuals are those from the Fe K fluorescence emission due to X-ray
reflection. 

The residuals that remain after fitting the continuum can be well fitted with a
distant reflection model component, in which the gas is assumed either to be
completely neutral or at a very low ionization stage, and no relativistic
effects are included.  We have tested this idea by implementing three different
(non-relativistic) reflection models, namely, \MYtorus, \Borus, and
\xillverCp, which we describe below.  The residuals of these fits are shown in
the last three panels of Figure~\ref{fig:appro1}, and the best-fit values
summarized in Table~\ref{tab:model1}.

{\it Model~1.1}: The \MYtorus\ reflection model \citep{mur09}
calculates the attenuation in the line-of-sight of X-rays produced by
a central source, together with the scattered continuum, and the
fluorescence emission from neutral iron and nickel, assuming a
toroidal geometry. In this model, the X-ray source emits a power-law continuum
with no cutoff at high energies. All elemental abundances are at their
Solar values. In our fit, all the parameters of the transmitted and
scattered components are tied to each other. The photon index is
linked to the one from the hot-corona component. The inclination is
fixed to 60$^\circ$, as it has no appreciable effect on the fit. Thus, the
column density and normalization are the only free parameters.

{\it Model~1.2}: The \Borus\ model \citep{bal18}, is similar to \MYtorus\ in
nature, but is more flexible as it provides additional tunable spectral
parameters such as the high-energy cutoff in the intrinsic continuum, the torus
covering factor, and the relative abundance of iron. The approximately toroidal
geometry assumed for the model employed here\footnote{We used table model 
{\tt borus02\_afe1p00\_v161220.fits} available at
\url{www.astro.caltech.edu/~mislavb/download}.} is the same as in the popular
model by \cite{bri11}, but the model is updated, expanded, and corrected for
known issues as described in \cite{liu15} and \cite{bal18}.  Like \MYtorus, \Borus\ allows us to
model the average column density of the torus separately from the line-of-sight
column density through the spectral shape of the reflection from material
outside of our line of sight.  As before, the inclination is degenerate in our
fits, which allows us to fix it at 60$^\circ$. Again, the photon index is
linked to the one describing the Comptonized emission of the hot corona.
Furthermore, the covering fraction was fixed to 50\%, and 
and the iron abundance is set to its Solar value.
Finally, the simple relation $E_\mathrm{cut}\sim (2-3)kT_\mathrm{e}$ is used
to link the cutoff at high energies with the electron temperature of the hot
corona \citep[e.g.;][]{pet01,gar15b}. While this is a crude approximation that
depends on the combination of temperature, optical depth, and geometry, we found that
it is adequate for this model fit. First, the value of $E_\mathrm{cut}$ is unconstrained
when set free to vary in the \Borus\ model, while all the other model parameters
remain unchanged. Furthermore, replacing the Comptonization continuum by a simple
cutoff power-law model provides an identical fit with $E_\mathrm{cut}=95-175$\,keV
(90\% confidence), consistent with $E_\mathrm{cut}\sim 3kT_\mathrm{e}$. Thus,
we use this relation to link the cutoff in the reflection model with the temperature
of the hot corona.
As in the case of Model~1.1, the only free parameters are the column density and
the normalization. 

%
\graphicspath{{/Users/javier/mrk509/march-2017/plots/}}

{\it Model~1.3}: We reproduced the observed residuals with our ionized
reflection model \xillverCp. This particular flavor of the model
computes the reflected spectrum using an illumination continuum
produced by the Comptonization model \nthComp. \nthComp\ is a more
physically consistent treatment than the standard and commonly used
power-law continuum with an exponential cutoff. While \xillverCp\ has
a more accurate treatment of the reflection by self-consistently
solving the ionization balance and radiative transfer, the geometrical
considerations are much more simplistic than in \MYtorus\ or
\Borus. In \xillverCp\ a single zone, plane-parallel slab is assumed.
Despite this approximation, this model also provides a satisfactory
fit to the data (Figure~\ref{fig:appro1}, bottom panel).  As before,
the slope of the illumination is fixed to that in the hot-corona
model.  Moreover, we fixed the ionization parameter, defined as the ratio
of the ionizing flux to the gas density ($\xi=4\pi F_\mathrm{x}/n_\mathrm{e}$)
to its minimum value in the model
order to mimic reflection off neutral gas ($\log\xi$/erg\,cm\,s$^{-1}=0$).
We assumed Solar abundance of iron. In this case, the inclination has a
small but noticeable effect in the fit, with the best-fit value pegged
at its maximum ($i=89^\circ$). Fixing the inclination to a more
reasonable value (e.g., $i=60^\circ$) worsens the fit significantly
($\Delta\chi^2\sim60$), due to strong residuals at high energies and
near the Fe K band. While this could be taken as the possible presence
of a broad Fe line component, its statistical significance is low.
Moreover, given the simplicity of the \xillverCp\  model in its
geometrical considerations, we do not interpret the derived inclination
as a meaningful estimate.
In addition to the inclination, the
normalization is the only other free parameter in this fit.

%
\graphicspath{{/Users/javier/mrk509/march-2017/plots/}}
\begin{figure*}[ht!]
\centering
\includegraphics[width=\linewidth,trim={0. 0. 0. 0.}]{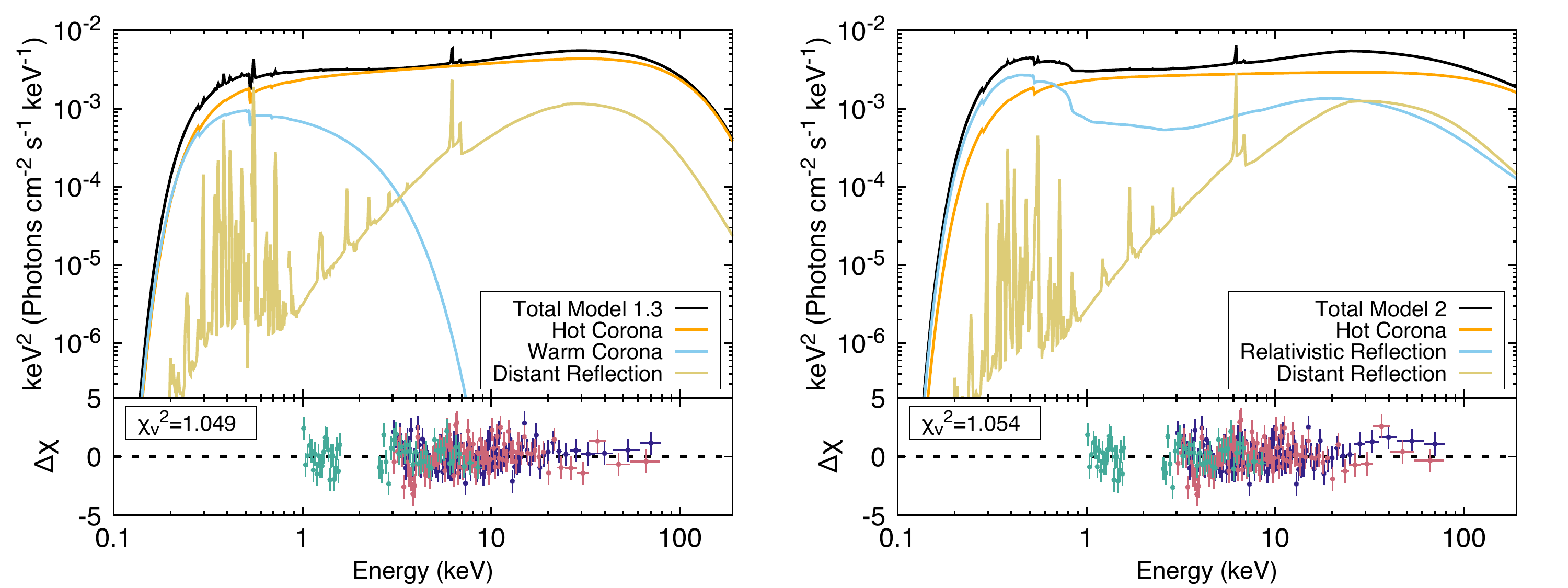}
\caption{Model components (top panels) and residuals (bottom panels) for the
best fits using the warm corona approach (Model~1.3, left) and the relativistic
reflection approach (Model~2, right) to describe the soft excess observed in
Mrk~509.}
\label{fig:appro2}
\end{figure*}

From a statistical point of view, these three models are
indistinguishable. Only very small differences in the goodness of the
fit are apparent in the bottom of Table~\ref{tab:model1}. From these,
the fit with the \Borus\ model (Model~1.2) is slightly worse, but
with a marginal increase in $\chi^2$ of $\sim 4$ when compared to
the other two. From the ratio plots shown in
Figure~\ref{fig:appro1}, it appears that these three models perform equally
well in describing the data.
Despite some small differences, these three fits
share the same relevant aspects. First, no inner-disk (relativistic)
reflection is required in any of the fits, as no significant residuals
remain in the Fe K region. Second, the electron temperature of the hot
corona is relatively low ($kT_\mathrm{e}^\mathrm{HC}\sim30$\,keV), which
suggests a low energy cutoff in the continuum.
Finally, the electron temperature of the warm corona is
similar in all the fits at $kT_\mathrm{e}^\mathrm{WC}\sim0.4-0.5$\,keV, which
is consistent with values previously derived by \cite{pet13}.


\def\nh{$4.25$}	   
\def\zz{$0.035$}   
\def\HCkTbb{$100$} 
\def\WCkTbb{$3$}   
\def\WCga{$2.5$}   

\def\mIHCga{$1.84 \pm 0.01$}   
\def\mIHCkTe{$26^{+6}_{-4}$}   
\def\mIHCnor{$1.33 \pm 0.02$}  
\def\mIWCkTe{$0.39 \pm 0.07$}  
\def\mIWCnor{$0.34 \pm 0.07$}  
\def\mIMYnh{$1.06 \pm 0.14$}   
\def\mIMYnorm{$1.25 \pm 0.12$} 
\def\mICB{$1.026 \pm 0.005$} 
\def\mICS{$0.95 \pm 0.04$}   
\def\mIchi{$1799.5$}
\def\mIdof{$1716$}
\def\mIredc{$1.049$}

\def\mIIHCga{$1.83 \pm 0.01$}   
\def\mIIHCkTe{$29^{+6}_{-4}$}   
\def\mIIHCnor{$1.32 \pm 0.02$}  
\def\mIIWCkTe{$0.40 \pm 0.07$}  
\def\mIIWCnor{$0.35 \pm 0.06$}  
\def\mIIBonh{$1.17 \pm 0.20$}   
\def\mIIBonorm{$0.93 \pm 0.08$} 
\def\mIICB{$1.026 \pm 0.005$} 
\def\mIICS{$0.95 \pm 0.03$}   
\def\mIIchi{$1804.0$}
\def\mIIdof{$1716$}
\def\mIIredc{$1.051$}

\def\mIIIHCga{$1.84 \pm 0.01$}   
\def\mIIIHCkTe{$29^{+6}_{-4}$}   
\def\mIIIHCnor{$1.29 \pm 0.02$}  
\def\mIIIWCkTe{$0.50^{+0.05}_{-0.08}$}  
\def\mIIIWCnor{$0.37 \pm 0.07$}  
\def\mIIIincl{$89$}            
\def\mIIIXinorm{$0.25^{+0.02}_{-0.01}$} 
\def\mIIICB{$1.026 \pm 0.005$} 
\def\mIIICS{$0.95 \pm 0.03$}   
\def\mIIIchi{$1800.4$}
\def\mIIIdof{$1716$}
\def\mIIIredc{$1.049$}

%
\begin{deluxetable*}{lccccccccc}[ht!]
\tablecaption{Best-fit parameters and fit statistics for the three models
featuring a warm corona prescription.
\label{tab:model1}}
\tablecolumns{6}
\tablewidth{0pt}
\tablehead{
\colhead{Description} & \colhead{Component} & \colhead{Parameter} &
\colhead{Model~1.1} & \colhead{Model~1.2} & \colhead{Model~1.3}
}
\startdata
Galactic Absorption & {\tt TBabs}         & $N_\mathrm{H}$ ($10^{20}$ cm$^{-2}$)& \nh\       &  \nh\       &  \nh\       \\
\hline
Hot Corona          & {\tt nthComp}       & $kT_\mathrm{BB}$ (eV)               & \HCkTbb\   &  \HCkTbb\   &  \HCkTbb\   \\
Redshift            & {\tt nthComp}       & $z$                                 & \zz\       &  \zz\       &  \zz\       \\
Warm Corona         & {\tt nthComp}       & $kT_\mathrm{BB}$ (eV)               & \WCkTbb\   &  \WCkTbb\   &  \WCkTbb\   \\
                    & {\tt nthComp}       & $\Gamma$                            & \WCga\     &  \WCga\     &  \WCga\     \\
\hline
Hot Corona          & {\tt nthComp}       & $\Gamma$                            & \mIHCga\   & \mIIHCga\   & \mIIIHCga\  \\
                    & {\tt nthComp}       & $kT_\mathrm{e}^\mathrm{HC}$ (keV)   & \mIHCkTe\  & \mIIHCkTe\  & \mIIIHCkTe\ \\
                    & {\tt nthComp}       & $N_\mathrm{HC}$ ($10^{-2}$)         & \mIHCnor\  & \mIIHCnor\  & \mIIIHCnor\ \\
Warm Corona         & {\tt nthComp}       & $kT_\mathrm{e}^\mathrm{WC}$ (keV)   & \mIWCkTe\  & \mIIWCkTe\  & \mIIIWCkTe\ \\
                    & {\tt nthComp}       & $N_\mathrm{WC}$ ($10^{-2}$)         & \mIWCnor\  & \mIIWCnor\  & \mIIIWCnor\ \\
\hline
Neutral Reflection  & {\tt MYtorus}       & $N_\mathrm{H}$ ($10^{24}$ cm$^{-2}$)& \mIMYnh\   & \nodata     & \nodata \\
                    & {\tt MYtorus}       & $N_\mathrm{MY}$\tablenotemark{b}                     & \mIMYnorm\ & \nodata     & \nodata \\
\hline
Neutral Reflection  & {\tt Borus02}       & $N_\mathrm{H}$ ($10^{24}$ cm$^{-2}$)& \nodata    & \mIIBonh\   & \nodata \\
                    & {\tt Borus02}       & $N_\mathrm{B}$\tablenotemark{b}                      & \nodata    & \mIIBonorm\ & \nodata \\
\hline
Neutral Reflection  & {\tt xillverCp}     & $i$ (deg)                           & \nodata    & \nodata     & \mIIIincl\tablenotemark{a} \\
                    & {\tt xillverCp}     & $N_\mathrm{Xi}$ ($10^{-2}$)\tablenotemark{b}         & \nodata    & \nodata     & \mIIIXinorm\ \\
\hline
Cross-normalization & \nustar\ FPMB       & $C_\mathrm{FPMB}$                   & \mICB\     & \mIICB\     & \mIIICB\  \\
                    & \suzaku\ XIS        & $C_\mathrm{XIS}$                    & \mICS\     & \mIICS\     & \mIIICS\  \\
\hline
                    & $\chi^2$            &                                     & \mIchi\    & \mIIchi\    & \mIIIchi\  \\
                    & $\nu$               &                                     & \mIdof\    & \mIIdof\    & \mIIIdof\  \\
                    & $\chi_{\nu}^2$      &                                     & \mIredc\   & \mIIredc\   & \mIIIredc\ \\
\enddata
\tablenotetext{a}{Parameter pegged at its maximum value.}
\tablenotetext{b}{Model normalizations in $10^{20}$\,photons\,cm$^{-2}$\,s$^{-1}$.}
\tablecomments{Model~1.1: {\tt TBabs*(nthComp+nthComp+MYtorus)}, Model~1.2:
{\tt TBabs*(nthComp+nthComp+borus02)}, and Model~1.3: {\tt
TBabs*(nthComp+nthComp+xillverCp)}. The parameters listed in the first block
were assumed fixed at the same value in all the models.
}
\end{deluxetable*}

\subsection{Approach 2: Fitting the Soft Excess With Relativistic Reflection}\label{sec:appro2}

Another approach that has been proposed in the past to explain the
soft excess in AGN is relativistic reflection
\citep[e.g.,][]{cru06,fab09,nard12,wal13}. As the X-rays from the central source
illuminate the inner regions of the accretion disk, the reflected or
reprocessed radiation displays a spectrum rich in fluorescence lines
and other atomic features. This spectrum is particularly populated
with emission lines in the low energy range ($\lesssim 1$\,keV), where
most of the K-shell transitions from low-$Z$ elements occur. As the
reprocessing is produced near the supermassive black hole,
relativistic effects will blur and skew all the atomic features,
effectively smoothing the entire reflected spectrum. As a result, this
component can in fact produce enough flux at low energies to explain
the observed soft excess. Furthermore, we have recently
shown that this effect is further
enhanced if the density of the reflecting material lies above the
typically assumed value of $n_\mathrm{e} = 10^{15}$\,cm$^{-3}$, due to
the extra heating produced by the increased free-free emission
\citep{gar16b}.

In order to test this approach, we replaced the warm-corona component
with a relativistic-reflection component. For this, we implemented our
model \relxillD, which describes both the incident Comptonized
continuum and the reflection spectra calculated with our code
\xillver\ \citep{gar10,gar13a} in the case of a high-density gas
\citep[\xillverD,][]{gar16b}, taking into account all the relativistic
effects \citep{dau13,gar14a}. While the \relxillD\ model has the
advantage of providing the gas density as a free parameter, one
limitation is that the illumination continuum assumed is a power-law
spectrum with an e-folded cutoff fixed at 300\,keV (instead of the
Comptonization continuum used in \xillverCp). 
However, freeing up the cutoff energy will only
introduce a significant effect in the fit if the curvature imprinted
in the power-law continuum falls within the covered bandpass and it
can be detected given the instrument's signal-to-noise.

In this fit, from here on Model~2, the distant
(non-relativistic) reflection is still modeled with \xillverCp, as in
Model~1.3. For the relativistic reflection, we use the specific
flavor of \relxillD, namely \relxilllpD, in which a lamppost geometry
is assumed for the hot corona \citep{dau13,dau16} that is
self-consistently linked with the reflected continuum. The slopes of
both the distant and inner-disk reflection components are tied to that
in the hot corona, as well as the electron temperature in
\xillverCp. The inclination of the system is tied among the two
reflection components. Unlike the previous fits with Models~1.1--1.3, in
this case the electron temperature of the hot corona is loosely constrained.
Fixing $kT_\mathrm{e}=30$\,keV (similar to the value found with the fits
in Section~\ref{sec:appro1}), results in a significantly worse fit (with $\chi^2$
increasing by $\sim 40$), and obvious residuals in excess at high energies. This
indicates that this particular fit prefers a cutoff a much larger energies.
Adopting once again the simple approximation $E_\mathrm{e}=3kT_\mathrm{e}$, we fixed 
the electron temperature of the hot corona at 100\,keV (i.e., one third of 
the cutoff energy of 300\,keV in the \relxilllpD\ component).  The best-fit parameters are
summarized in Table~\ref{tab:model2}.

In terms of fit statistics, the relativistic reflection prescription
reproduces the data similarly well as the warm corona prescription from
Models~1. The fit with Model~2 is marginally worse, with an increase of
$\Delta\chi^2\sim 6-10$, despite using three more free parameters. It is,
however, unclear if any of these fits is preferred on statistical grounds. The
model components and residuals of the fits with the two scenarios (Models~1.3
and 2) are compared in Figure~\ref{fig:appro2}. The two models are almost
identical in the band covered by the data, with the largest differences
occurring around $30-60$\,keV for Model~2. These residuals are possibly due to
the fact that the reflection model used here was calculated using an e-folded
power-law illumination spectrum with a high-energy cutoff fixed at 300\,keV,
rather than a proper Comptonization continuum.  On the other hand, we also note
that Model~2 allows for a softer continuum ($\Gamma=1.96$) than Model~1.3
($\Gamma=1.84$), which can also affect the way the model fits the rollover
at high energies.

\def\mHCkTe{$100$}                    
\def\mHCga{$1.96^{+0.01}_{-0.03}$}    
\def\mHCnor{$1.30^{+0.01}_{-0.03}$}   
\def\mlph{$1.53^{+0.01}_{-0.25}$}     
\def\mlpa{$>0.993$}                   
\def\mlpi{$69.7^{+2.8}_{-1.7}$}       
\def\mlplxi{$2.31^{+0.39}_{-0.15}$}   
\def\mlpafe{$<0.82$}                  
\def\mlplne{$>18.2$}                     
\def\mlpnor{$2.23^{+9.95}_{-0.44}$}   
\def\mXinor{$2.02^{+0.34}_{-0.13}$}   
\def\mCB{$1.026 \pm 0.005$} 
\def\mCS{$0.95 \pm 0.02$}   
\def\mchi{$1806.1$}
\def\mdof{$1713$}
\def\mredc{$1.054$}

Despite its statistical match, the relativistic reflection component
(\relxilllpD) requires extreme parameters, i.e., low coronal height
($h=$\mlph\,$R_\mathrm{Hor}$), and close to maximum spin ($a_*$\mlpa), together
with a large gas density ($\log n_\mathrm{e}$/cm$^{-3}$ \mlplne). This
configuration results in a soft and featureless spectrum, with a strong broad
emission at low energies, which is required to fit the soft excess.

%
\begin{deluxetable*}{lccc}[ht!]
\tablecaption{Best-fit parameters and fit statistic for the model featuring relativistic reflection.
\label{tab:model2}}
\tablecolumns{4}
\tablewidth{0pt}
\tablehead{
\colhead{Description} & \colhead{Component} & \colhead{Parameter} & \colhead{Model~2}
}
\startdata
Galactic Absorption & {\tt TBabs}         & $N_\mathrm{H}$ ($10^{20}$ cm$^{-2}$) & \nh\tablenotemark{a} \\
\hline
Hot Corona          & {\tt nthComp}       & $kT_\mathrm{BB}$ (eV)                & \HCkTbb\tablenotemark{a}  \\
                    & {\tt nthComp}       & $z$                                  & \zz\tablenotemark{a}      \\
                    & {\tt nthComp}       & $\Gamma$                             & \mHCga\   \\
                    & {\tt nthComp}       & $kT_\mathrm{e}^\mathrm{HC}$ (keV)    & \mHCkTe\tablenotemark{a}  \\
                    & {\tt nthComp}       & $N_\mathrm{HC}$ ($10^{-2}$)          & \mHCnor\  \\
\hline
Relativistic Reflection & {\tt relxilllpD}& $h$ ($R_\mathrm{Hor}$)               & \mlph\    \\
                    & {\tt relxilllpD}    & $a_*$ ($cJ/GM^2$)                    & \mlpa\    \\
                    & {\tt relxilllpD}    & $i$ (deg)                            & \mlpi\    \\
                    & {\tt relxilllpD}    & $\log\xi$ (erg\,cm\,s$^{-1}$)        & \mlplxi\  \\
                    & {\tt relxilllpD}    & $A_\mathrm{Fe}$ (Solar)              & \mlpafe\  \\
                    & {\tt relxilllpD}    & $\log n_\mathrm{e}$ (cm$^{-3}$)      & \mlplne   \\
                    & {\tt relxilllpD}    & $N_\mathrm{r}$ ($10^{-2}$)\tablenotemark{b}           & \mlpnor\  \\
Distant Reflection  & {\tt xillverCp}     & $N_\mathrm{x}$ ($10^{-4}$)\tablenotemark{b}           & \mXinor\  \\
\hline
Cross-normalization & \nustar\ FPMB       & $C_\mathrm{FPMB}$                    & \mCB\    \\
                    & \suzaku\ XIS        & $C_\mathrm{XIS}$                     & \mCS\    \\
\hline
                    & Flux (erg\,cm$^{-2}$\,s$^{-1}$)   & 2--10\,keV            & $4.6\times 10^{-11}$ \\
                    &                                   & 20--40\,keV           & $3\times 10^{-11}$ \\
\hline
                    & $\chi^2$            &                                      & \mchi\   \\
                    & $\nu$               &                                      & \mdof\   \\
                    & $\chi_{\nu}^2$      &                                      & \mredc\  \\
\enddata
\tablenotetext{a}{Parameter fixed to the quoted value.}
\tablenotetext{b}{Model normalizations in $10^{20}$\,photons\,cm$^{-2}$\,s$^{-1}$.}
\tablecomments{Model~2: {\tt TBabs*(nthComp+relxillD+xillverCp)}.}
\end{deluxetable*}

\section{Discussion}\label{sec:disc}

In the previous section we presented several model fits to the observational
data of \mrk509. These models are based on two different scenarios to explain
the origin of the soft excess in the spectrum: the warm corona and the relativistic
reflection picture. In either case, strong signatures of reflection are observed 
(i.e., Fe K emission and K edge, plus a Compton hump). This signal is consistent 
with low-ionization reflection from a structure located at a farther distance such that no relativistic
effects are observed. Models for Compton-thick AGN (\MYtorus\ and \Borus; Models~1.1 and 1.2),
and nearly-neutral reflection
from a single plane-parallel slab (\xillverCp; Model~1.3) all provide
equally good fits to the data. This implies that the geometrical considerations 
for the distribution of gas in the line-of-sight are relatively unimportant.
Moreover, we notice that no other components are required to fit the Fe K emission,
while \cite{pon13} reported both a narrow ($\sigma=0.027$\,keV), plus a resolved 
($\sigma=0.22$\,keV) Gaussian feature for the Fe K line in their analysis of previous
\chandra\ grating data. However, these two components were unresolved in their
\xmm\ and \suzaku\ data, and likewise are expected to be unresolved in our \xmm\ and 
\nustar\ data. This is possibly the reason why
Models~1.1--1.3 are able to reproduce the spectral features without any additional components.
Weak ionized emission features were also reported by \cite{pon13}, which could be
attributed to Fe~{\sc xxv-xxvi}. We do not find evidence for these additional components,
possibly due to the lower signal-to-noise of our data.

For the sake of comparison, we will now focus on the fits performed with Models~1.3 (warm
corona) and Model~2 (relativistic reflection at high densities), and discuss
the physical implications of each scenario.

\subsection{Implications of the Warm Corona Model}\label{sec:wcdisc}

In the warm corona model, the soft emission observed in excess of the
hard power-law continuum originates in Comptonization of thermal
disk photons into a warm ($T\sim0.5-1$\,keV or $\sim 0.6-1.2\times
10^{7}\,$K) and optically thick ($\tau_\mathrm{T}\sim 10-20$) corona
\citep{wal93,mag98,don12}. This warm corona has been described as a
slab sitting on top of a passive accretion disk covering roughly
10--20\,$R_\mathrm{g}$ of the inner region \citep[e.g.,][]{pet13}.
One argument that favors this scenario is the observed correlation
between the optical-UV and the soft X-ray emission \citep{meh11}. As
shown in our fits to Models~1.1-1.3, the warm corona model provides a
satisfactory description of the data in combination with a distant
reflection component, without the requirement of relativistic
reflection. 

In this case, the temperature of the hot corona (the one
responsible for the hard power-law continuum) is found to be relatively low
($kT_\mathrm{e} \sim 25-35$\,keV or $\sim 3-4\times 10^{8}\,$K).
While low coronal temperatures were not common in earlier studies of AGN
\citep[e.g.,][]{mar16}, several recent \nustar\ measurements have reported
relatively cold coronae, namely $\sim 50$\,keV
\citep[IC~4329A,][]{bre14}, $\sim 25$\,keV \citep[MCG$-$05-23-016,][]{bal15}, 
$\sim 40$\,keV \citep[NGC~5548,][]{urs15},
$\sim 12$\,keV \citep[GRS~1734$-$292,][]{tor17}, $\sim 35$\,keV \citep[IRAS~05189$-$2524,][]{xu17}, 
and $\sim 15$\,keV \citep[Ark~564,][]{kar17}. Moreover, \cite{ric17} have
also reported a handful of sources with low cutoff energies fitting e-folded power-law
models to sources from the \swift/BAT sample, and found that those sources appear to be the ones
with the highest Eddington ratios. Meanwhile, \cite{tor18} have reported more
reliable coronal temperatures for a sample of AGN by implementing thermal
Comptonization models, in which most of the sources are found to have
coronal temperatures below $\sim 60$\,keV.

While the warm-corona model has been successfully used in several
other sources \citep[see][and references therein]{pet18}, its physical
origin and implications have yet to be fully
explained. \cite{cze03} argued that a warm Comptonizing skin on top
the accretion disk under radiation pressure instabilities could
explain the observed X-ray spectra from quasars and narrow line Seyfert
AGN. \cite{roz15} investigated the properties of such a corona by solving
the radiative transfer for a grey atmosphere. More recently,
\cite{pet18} presented a theoretical discussion to explain the warm
corona based on simple photon conservation arguments, concluding that
most of the energy dissipation takes place in the warm corona rather
than in the accretion disk. Meanwhile, \cite{kau18} proposed that bulk
Comptonization from turbulence due to magneto-rotational instabilities
can explain the warm corona. These authors argue, however, that this picture is only applicable
to systems with high accretion rate, possibly of the order or larger
than the Eddington limit. Crucially, all these theoretical studies
share the same fundamental limitation: they neglect the effects of
atomic photoelectric absorption, which is likely to be a dominant
process in optically-thick atmospheres.

%
\graphicspath{{/Users/javier/mrk509/march-2017/plots/}}
\begin{figure*}
\centering
\includegraphics[width=\linewidth,trim={0. 0. 0. 0.}]{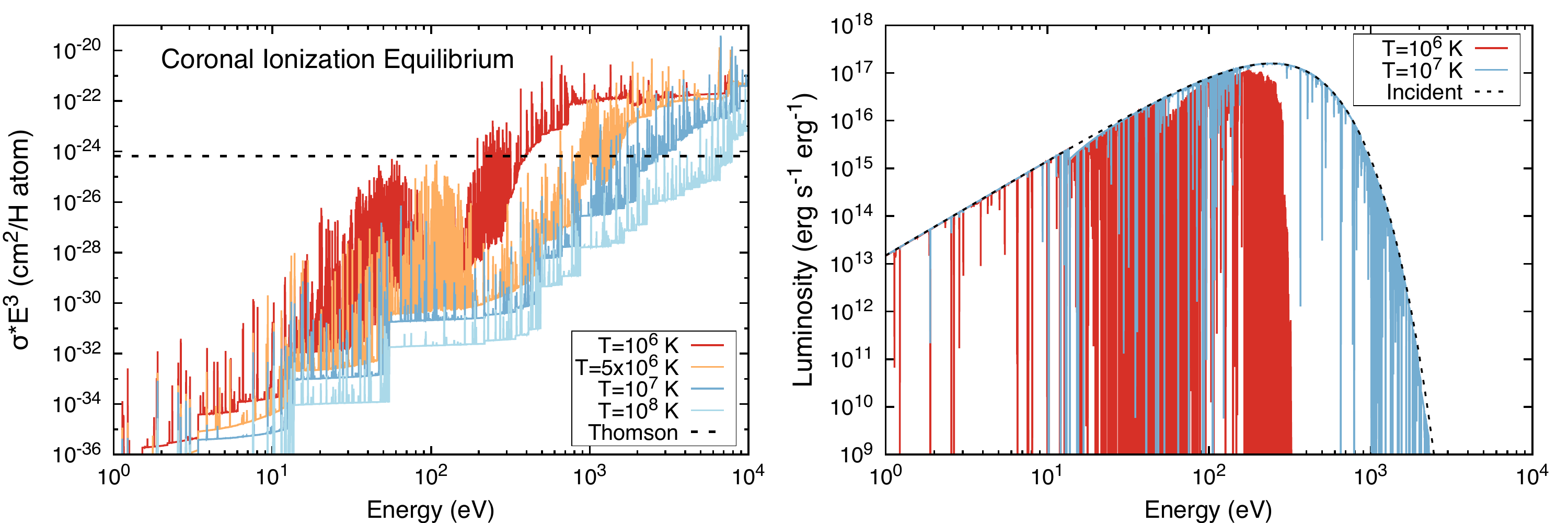}
\caption{Calculations for a gas under coronal ionization equilibrium for
different temperatures (as indicated), and the parameters that describe
a warm corona: $n_\mathrm{e}=10^{12}$\,cm$^{-3}$, $\tau_\mathrm{T}=6.65$, and 
$L=10^{46}$\,erg\,s$^{-1}$. The left panel shows the photoelectric 
opacity as a function of energy, compared to the Thomson electron opacity.
The right panel shows the resulting transmitted spectra. For clarity, 
only two cases are displayed, those for the lowest and highest temperatures 
derived from fitting the warm corona model to observational data. 
The incident blackbody spectrum for the $T=10^7$\,K is also shown.
}
\label{fig:cie}
\end{figure*}

It is interesting to describe the basic properties of the warm corona based on
the average quantities obtained from fits to observational data \citep[e.g.,
$\tau_\mathrm{T}\sim10$ and $kT_\mathrm{e}\sim 0.5$\,keV,][]{pet18}.
The vertical extension of this
corona can be estimated as
\begin{equation}\label{eq:thickness}
z=\tau_\mathrm{T}/(\sigma_\mathrm{T}n_\mathrm{e}),
\end{equation}
where $\sigma_T=6.65\times10^{-25}$\,cm$^2$ is the Thomson cross section and
$n_\mathrm{e}$ is the electron density. Therefore $z\sim1.5\times10^{24} \tau_\mathrm{T} n_\mathrm{e}^{-1}$\,cm,
or in units of the gravitational radius $R_g=GM/c^2\approx 1.5\times 10^{13} (M_8/\msun)$\,cm,
\begin{equation}
z/{R}_g \sim 10^{11} (M_8/\msun) (\tau_\mathrm{T}/n_\mathrm{e}),
\end{equation}
where $G$ is the gravitational constant, $c$ is the speed of light, and $M_8=10^8\msun$.
In the case of Mrk~509, $M_8/\msun\sim1$ \citep{pet04},
and thus the density must be of the order of $n_\mathrm{e} \sim
10^{12}$\,cm$^{-3}$ or higher for the warm corona to have a reasonable
($z\sim R_g$) geometrical thickness. Moreover, for sources with $M\sim10^5-10^6 \msun$, this
estimate implies densities for the warm corona of the order of the 
typical values used for the accretion disk atmosphere in X-ray reflection
calculations \citep[e.g., $n_\mathrm{e}\sim10^{15}$\,cm$^{-3}$;][]{ros05,gar10}.

One requirement for the warm corona scenario is to ensure that electron
scattering is the dominant source of opacity.  However, \cite{kro84} showed
that for an optically-thin gas under coronal ionization equilibrium (CIE), the
photoelectric opacity dominates the soft band for $T\sim10^6$\,K, and even at
$T\sim 10^7$\,K it is comparable to the Thomson opacity at $\sim 1$\,keV (see
their Fig.~1). We have tested this argument by computing simulations for an
optically-thick plasma under CIE using the latest version of the {\sc xstar}
code \citep{kal01}, with the appropriate parameters that describe a warm
corona: $n_\mathrm{e}=10^{12}$\,cm$^{-3}$, $\tau_\mathrm{T}=6.65$ (corresponding to the
maximum column allowed by the model, $N_\mathrm{H}=10^{25}$\,cm$^{-2}$),
$L_\mathrm{x}=10^{46}$\,erg\,s$^{-1}$ (which is in fact larger than the value typically
measured for this source), and cosmic abundances. The incident spectrum is assumed
to be a blackbody at the given gas temperature. Figure~\ref{fig:cie} (left)
shows the resulting photoelectric opacity as function of energy for different
gas temperatures, in comparison with the Thomson opacity for electron
scattering $\sigma_\mathrm{T}$. This demonstrates that even in the
optically-thick case, photoelectric opacity dominates over a wide range of
energies, particularly around or above 1\,keV, for the range of temperatures required by
the warm corona, i.e., $kT\sim 0.1 - 1$\,keV ($T\sim 10^6 - 10^7$\,K).  The right
panel in Figure~\ref{fig:cie} shows the transmitted spectra for these two CEI
calculations. At $T=10^6$\,K the original disk blackbody emission is heavily
absorbed and modified, with strong photo-absorption at almost all energies and with
no emission above $\sim 300$\,eV.  The
situation is better at $T=10^7$\,K, although strong absorption is still present,
particularly around 0.1\,keV and 1\,keV. We found in general that for electron
scattering to be a dominant source of opacity, temperatures well above $10^7$\,K
are required.


Another possibility is to instead invoke a gas under photoionization
equilibrium, since a radiation field strong enough can be responsible for
stripping most of the ions and thus considerably reducing the total
photoelectric opacity. This is in fact relevant since one expects the ionization
of the warm corona to be fairly large, from simple arguments.  We start by
using the standard definition of the ionization parameter $\xi=L/(n_\mathrm{e}R^2)$,
where $L$ is the luminosity and $R$ is the distance from a generic source of 
radiation (e.g., the hot corona) to the warm
corona. For a thin disk, $z/R=$constant$\sim0.1$, and using
Equation~\ref{eq:thickness}
\begin{equation}
\xi = 10^{-1}\frac{\sigma_\mathrm{T}L}{\tau_\mathrm{T}R}
\end{equation}
or
\begin{equation}\label{eq:xi1}
\xi~[\mathrm{erg\,cm\,s}^{-1}] \sim 10^7 (L/L_\mathrm{Edd})(R/R_g)^{-1},
\end{equation}
where $L_\mathrm{Edd}=1.26\times10^{46}(M_8/\msun)$\,erg\,s$^{-1}$ is the
Eddington luminosity. So for $L/L_\mathrm{Edd}=0.1$ and $R=10R_g$,
$\xi\sim10^5$\,erg\,cm\,s$^{-1}$. A different estimate can be made if the
ionization is assumed to be due to the thermal emission from the accretion
disk. Using the definition of the ionization parameter $\xi = 4\pi F/n_\mathrm{e}$,
and the local flux from the disk
\begin{equation}
F_d = \frac{3\pi}{8} \frac{GM\dot{M}}{R^3}
\end{equation}
and $\dot{M} = L/(\eta c^2)$, we find
\begin{equation}
\xi = \frac{0.15\pi^2\sigma_\mathrm{T}G}{\tau_\mathrm{T}\eta c^2}\frac{ML}{R^2},
\end{equation}
where $\eta\approx0.1$ is the accretion efficiency.
This last equation can be rewritten as
\begin{equation}\label{eq:xi2}
\xi~[\mathrm{erg\,cm\,s}^{-1}] \sim 10^9 (L/L_\mathrm{Edd})(R/R_g)^{-2},
\end{equation}
and thus for $L/L_\mathrm{Edd}=0.1$ at $R=10 R_g$ we get $\xi\sim
10^6$\,erg\,cm\,s$^{-1}$.  While this expression results in a larger ionization
than the estimate in Equation~\ref{eq:xi1}, it decreases quadratically (rather
than linearly) with radius. It is also interesting that both expressions are
independent of the black hole mass.

%
\graphicspath{{/Users/javier/mrk509/march-2017/plots/}}
\begin{figure*}
\centering
\includegraphics[width=\linewidth,trim={0. 0. 0. 0.}]{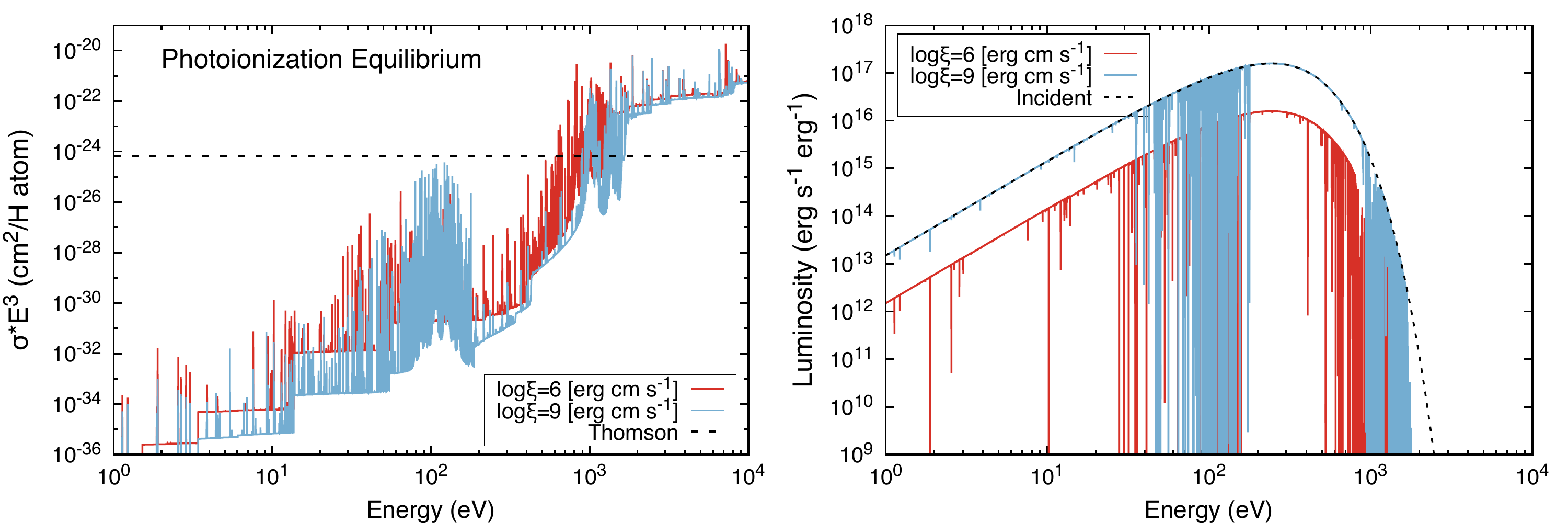}
\caption{Calculations similar to those in Figure~\ref{fig:cie}, but for the case
of a gas under photoionization equilibrium, at two different ionizations:
$\xi=10^6$ and $10^9$\,erg\,cm\,s$^{-1}$. As before, left panel shows the
photoelectric opacity as a function of energy together with the Thomson
opacity, while the right panel shows the resulting transmitted spectra. Even at
the highest ionization predicted by our estimates of the warm corona, the
original blackbody spectrum (dashed line) is severely modified by photoelectric absorption.
}
\label{fig:pie}
\end{figure*}

Although Equations~\ref{eq:xi1} and \ref{eq:xi2} predict fairly large
ionization for the warm corona, this is only true for the case of an
optically-thin slab.  For large optical depths ($\tau_T \simeq 1$), the
ionization will quickly decrease in the deeper regions of the gas, and
photoelectric absorption can be as important or more than the Thomson
opacity. These results are generally in line with the seminal calculations
presented by \cite{ros78}, where they considered the photoionization of
isothermal spheres at $T=10^6-10^7$\,K, with $\tau_\mathrm{T}=6$, and
$n_\mathrm{e}=10^{16}$\,cm$^{-3}$.  They found that despite the very high
ionization at the center of the cloud, in the outer parts ions such as Fe~{\sc xxii}
were still dominant, producing distinct spectral features.

As before, we use the {\sc xstar} code to test this scenario by producing the
solution for a plasma under photoionization equilibrium (PIE) using the estimates
shown above; i.e., $\xi\sim 10^6$\,erg\,cm\,s$^{-1}$, $n_\mathrm{e} = 10^{12}$\,cm$^{-3}$,
and $N_\mathrm{H}=10^{25}$\,cm$^2$.  Using a blackbody with
$kT=0.1$\,keV as the input spectrum, the resulting gas temperature is $T\sim10^6$\,K. Despite the
large ionization, the photoelectric opacity near 1\,keV is still dominant (or at
least comparable) to the Thompson opacity (left panel in Figure~\ref{fig:pie}).
We repeated this calculation by raising the ionization to the largest value
predicted by Equation~\ref{eq:xi2} (i.e., $\xi\sim 10^9$\,erg\,cm\,s$^{-1}$), but
the net effect is small in reducing the photoelectric opacity. Just as in the
case of CIE, the transmitted spectra show strong absorption features in the
observable bandpass (right panel in Figure~\ref{fig:pie}). Despite the large
ionizing flux, the input spectrum is too soft to fully ionize the metals in the
gas.

None of the spectra resulting from either the CIE or PIE simulations are likely
to resemble the apparently featureless broad component required to fit the
soft excess. In the case of PIE, a harder spectrum extending to high energies
is likely to provide enough photons to fully ionize the medium, such as that
provided by the hot corona. However, strong photoionization will raise the
temperature and can only fully ionize the atmosphere if the optical depth
is much smaller than that inferred from fitting the warm corona model.
Moreover, strong
illumination of an optically-thick medium is expected to produce strong
reprocessing of the photons, which is a situation that closely resembles the 
relativistic reflection model. This alternative scenario is discussed in the 
next Section.

\subsection{Implications of the Relativistic Reflection Model}\label{sec:rrdisc}

The relativistic reflection model has also been proposed as a possible
explanation for the soft excess in AGN. When strong radiation is
produced in the central region close the black hole, the reprocessing
of the hard X-rays in the optically thick and relatively cold
accretion disk is an expected consequence. If the 
reflection occurs close enough to the horizon, the relativistic
effects will distort the spectrum, broadening and skewing all the
spectral features. Below $\sim 1$\,keV, a rich forest of fluorescence
emission lines produced by ions with nuclear charge lower than iron is
predicted \citep[e.g.,][]{ros05,gar10}. When the gravitational blurring
is extreme, these features will blend creating a single broad and
smooth excess at soft energies. When facing the difficulties in making
physical sense out of a featureless and broad spectrum emitted from a warm
corona, a relativistically blurred reflection spectrum provides an alternative
and somewhat more consistent interpretation. However, some caveats must
also be considered when adopting this model. Below we discuss this scenario
to explain the soft excess in \mrk509.

In their analysis of a sample of 25 ``bare'' AGN with {\it
Suzaku}, \cite{wal13} fitted $\sim 90$\,ks XIS/PIN spectra of Mrk\,509 using
a model consisting of ionized and relativistic plus neutral and distant
reflection components. Both components were modeled with {\tt reflionx}
\citep{ros05}.  A warm absorber component was also included and modeled with
{\sc xstar}.  The relativistic blurring applied to the ionized reflection
employed {\tt relconv} \citep{dau13}.  Two sets of fits were performed: one
with a fixed cross-normalization constant between the hard X-ray PIN and the
soft X-ray XIS detectors, and another in which this cross-normalization
$C_\mathrm{PIN/XIS}$ was allowed to vary.  The uncertainty introduced by
$C_\mathrm{PIN/XIS}$ has critical impact on the results for Mrk~509. In short,
two vastly different pictures emerge from the fits, simply due to differences
in the hard X-ray component.  In the first instance of fixed
$C_\mathrm{PIN/XIS}=1.17$, the system demands a high spin $a=0.86\pm 0.02$ and
face-on orientation (inclination of $i<18^{\circ}$). However, when
$C_\mathrm{PIN/XIS}$ is freed, it becomes loosely constrained
($C_\mathrm{PIN/XIS}<1.06$), while spin and inclination are drastically
affected: $a=0.36\pm 0.3$ and $i=50\pm5^{\circ}$. This is because the hard
X-ray band is essential for disentangling the power-law continuum from the
ionized reflection component, which motivated the \nustar\ observations
presented here.

The fit described in Section~\ref{sec:appro2} and shown in
Figure~\ref{fig:appro2} demonstrates that the relativistic reflection scenario
(Model~2) provides a good description of the present \suzaku\ and \nustar\
data for \mrk509, with results that are broadly consistent with the high-spin fits presented by
\cite{wal13}. Moreover, our fits have been carried out with updated reflection
models, which include more complete atomic data, improved radiative transfer
calculations, and the possibility for higher densities in the reflector. This
latter improvement is important to better describe the soft excess observed
below 1\,keV. 

It is worth noticing the relevance of the \nustar\ data in providing
high signal-to-noise data at hard energies, where most of the reflection signatures
are observed. This is particularly important because not only our \suzaku\ exposure
is shorter than that analyzed by \cite{wal13}, but also because our data lack the
high-energy coverage previously provided by the PIN instrument (not longer operational
in the last \suzaku\ cycle). In the case of the relativistic reflection Model~2, 
fitting the \suzaku\ data alone yield poor constraints to important parameters such
as spin ($a_*>0.702$), coronal height ($h=2.01^{+0.29}_{-2.74}$\,$R_\mathrm{Hor}$), 
and inclination ($i=51.4\pm 8^\circ$). Unsurprisingly, the disk density
is determined with a similar uncertainty ($\log n_\mathrm{e}$/cm$^{-3}>18$),
as this parameter is mostly sensitive to the soft-energy data.

\begin{figure}
\centering
\includegraphics[width=\linewidth,trim={10 0 10 0}]{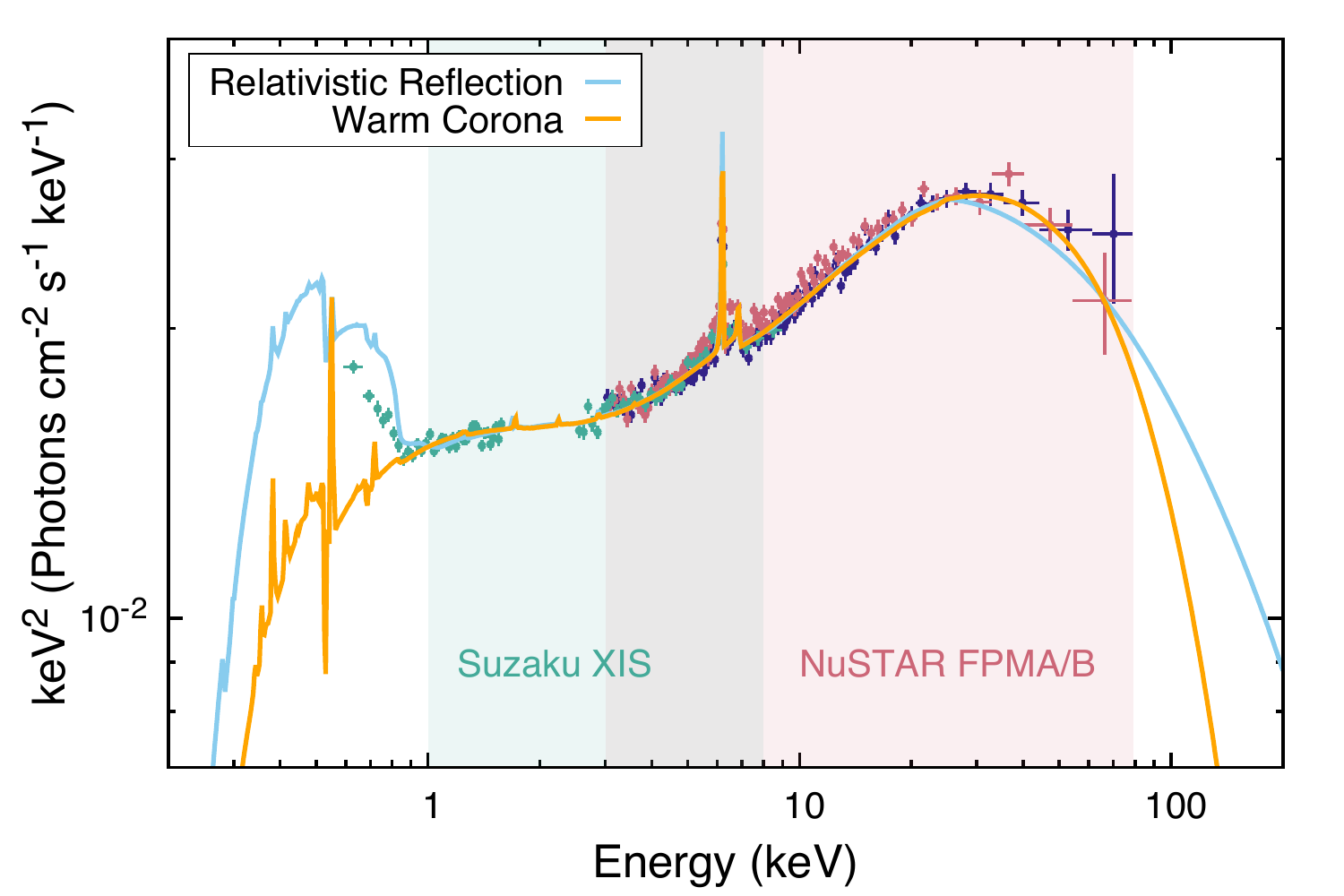}
\caption{Unfolded spectra of \mrk509\ as seen by \suzaku\ XIS and \nustar\ FPMA
and FPMB (data points), together with the two different scenarios for the soft excess 
(solid lines), the relativistic reflection (Model~2), and the warm corona (Model~1.3).
The shaded regions show the data from each instrument that was included in the fits.
}
\label{fig:models}
\end{figure}

When applied to both the \suzaku\ and \nustar\ data, the goodness-of-fit for
the relativistic reflection model ($\chi_{\nu}^2=1.054$) is very similar to
that from the fits with the warm corona picture ($\chi_{\nu}^2=1.049$,
Model~1.3). The similarity between the warm corona and the relativistic
reflection model has also been previously discussed by \cite{boi14}.  In
Figure~\ref{fig:models} we show these two models overplotted with the observed
data. It is clear that the two models are almost identical in the energy band
considered for the fits (1--79\,keV), which is shown with the shaded regions.
We emphasize that data below 1\,keV was excluded given concerns in the
calibration of \suzaku's instruments in this band towards the end the mission
(see Section~\ref{sec:data}).  We note, however, that when these data are
included (without refitting), they seem to favor the trend predicted by the
relativistic reflection model. Nevertheless, the lack reliable data below
1\,keV limits the analysis of the present study, as we cannot fully constrain
the overall shape of the soft excess. Thus, future observations with sensitive
coverage of both the soft and hard energy bands will become crucial to further
understand the nature of the soft excess in \mrk509, and several other AGN.

The small differences between the two
models seen at high energies ($\sim 30-50\,$keV, Figure~\ref{fig:models}), are
likely due to the fact that the reflection models used here were calculated
with a cutoff energy fixed at 300\,keV, while in the warm corona fit this
parameter is allowed to vary freely. This suggests that a lower coronal
temperature would be possible with the reflection model, but it is probably not
very well constrained, as it does not seem to affect the fit statistics
significantly.

The relativistic reflection model requires a large value for the black
hole spin (consistent with its maximum value, $a_*$\mlpa), and low coronal
height ($h=$\mlph\,$R_\mathrm{Hor}$).  While high spins and compact coronae are
commonly reported for AGN, a corona placed so close to the black hole implies a
very extreme configuration in which most of the radiation is focused toward the
disk due to the strong light bending \citep{dau16}. This configuration predicts
a reflection-dominated spectrum, different to the fit achieved with Model~2
(Figure~\ref{fig:appro2}, right).
Nonetheless,
modeling the primary source of X-rays as a point source in the rotational axis
is a rather simple and idealized description, and thus the derived parameters
need to be interpreted with care.  

The iron abundance is found to be close to its Solar value ($A_\mathrm{Fe}$\mlpafe).
Fixing $A_\mathrm{Fe}=1$ worsens the fit by $\Delta\chi^2\sim12$, having
no obvious effect on the rest of the model parameters. While Solar abundances
are the canonical expectation, much larger Fe abundances are commonly derived
from reflection modeling \citep{gar18}. However, recent studies indicate that high-density
reflection models (like the ones used here) lead to abundances closer to Solar 
\citep[e.g.;][]{tom18,jia18}, which is consistent with our findings. Moreover, visual inspection
of the residuals reveals no obvious signs of iron emission lines after the distant
reflection is accounted for (e.g., see Figure~\ref{fig:appro1}), suggesting that the reflection spectrum is primarily constrained by fitting the soft-excess.

The large density of the accretion disk derived from our fits ($\log n_\mathrm{e}$/cm$^{-3}$\mlplne)
also places this source in a somewhat extreme configuration.
For instance, \cite{sve94} derived analytic expressions for a hot corona around
a cold $\alpha$-disk system. Using their expression for the disk density in the
 radiation-pressure-dominated case (i.e., their Equation~8)
\begin{equation}
n_\mathrm{e} = \frac{1}{\sigma_\mathrm{T}R_\mathrm{S}} \frac{256\sqrt{2}}{27}
\alpha^{-1} r^{3/2} \dot{m}^{-2} \left[ 1-(3/r)\right]^{-1} (1-f)^{-3}.
\end{equation}
where $\alpha\approx 0.1$ is the standard \cite{sha73} dimensionless parameter
connecting the viscosity with the gas pressure, $r=10$ is the radius in units
of $R_\mathrm{S} = 2R_g$, $\dot{m}=0.1$, $n_\mathrm{e}=10^{19}$\,cm$^{-3}$, and
$f$ is the fraction of the total accretion power dissipated by the corona.  We
find $f=0.86$, which means that most of the accretion power
needs to be dissipated in the hot corona.  We note that more conservative
values can be found in the literature.  For example, \cite{vas07} reported
$f\sim0.11-0.45$ for a sample of 54 AGN.  Nevertheless, our estimate for
\mrk509, albeit extreme, is allowed within the applicability regime of the hot
corona and cold disk model. 

Meanwhile, high-density reflection models like the one used here have recently been used to
successfully describe the spectrum of the AGN IRAS~13224$-$3809
\citep{par17,jia18}, and Mrk~1044 \citep{mal18}, as well as of the black hole
binary Cyg~X-1 \citep{tom18}. In all these cases, fitting the observed soft
excess results in a lower (and more physical) iron abundance in the reflector
\citep[see also discussion in][]{par18}. However, in the case of Ark~120,
\cite{por18} find that the warm corona model provides a better description of
the data over the relativistic reflection picture, even when high-density
models were tested. In a multi-epoch study of Mrk~335, \cite{kee16} showed
that after fitting reflection above 3\,keV, a constant soft excess appears to
remain that is constant to the flux of the source. However, they have only used
standard relativistic reflection, as the high-density reflection models like the
ones used here were not available at the time.

%
\graphicspath{{/Users/javier/mrk509/march-2017/plots/}}

One argument against the relativistic reflection scenario (and consequently in
favor of the warm corona picture), on the other hand, is the apparent
discrepancy in the correlation between the strengths of the reflection
($R_\mathrm{f}$) and the soft excess ($R_\mathrm{SE}$) components predicted by
relativistic reflection models and that observed in Seyfert AGN.  \cite{boi16}
showed that while simulations with reflection models predict a positive
correlation between $R_\mathrm{f}$ and $R_\mathrm{SE}$ \citep[see
also][]{vas14}, observations of a sample of 42 AGN show a negative correlation.
They argued that this discrepancy can be overcome if instead the soft excess is
modeled with warm Comptonization models. However, their sample includes data
that are not simultaneous, which is likely to bias their results for sources
with strong variability. More importantly, their fits implement very simplistic
models for reflection, which are fundamentally incorrect to properly describe
the combination of distant (non-relativistic) and local (relativistic)
reflection. In many un-obscured AGN, the narrow (unblurred) reflection
component dominates the relativistic reflection signal \citep[e.g.,][]{ric14}. Thus, the reflection
fraction measured by \cite{boi16} is likely biased towards the strength of the
distant reflector. In this case, the anti-correlation with the strength of the
soft excess can be simply explained by geometrical effects.  For sources that
are more obscured, the emission from the inner-most regions will tend to be
reduced, which reduces the direct continuum (increasing $R_\mathrm{f}$) as well
as the local relativistic reflection component (decreasing $R_\mathrm{SE}$).

\section{Conclusions}\label{sec:concl}

We have presented an analysis of the X-ray spectrum (1--79\,keV) of the bright
Seyfert 1 AGN \mrk509. These data, obtained during April--June 2015 with
\suzaku\ and \nustar\, reveal signatures of X-ray reprocessing from an
optically thick and relatively cold material, a power-law continuum, and a
strong soft excess.  By performing fits of different modern models, we have
shown that these data can be described by a hot corona which produces the
power-law continuum (modeled with a standard Comptonization model), and a distant
reflection from a cold material (which can be described with a variety of
reflection models). Meanwhile, the soft excess can be fitted with either a warm
Comptonizing corona, or with a relativistically blurred high-density reflection
model.  These two prescriptions imply two very different interpretations of the
observed spectrum, and they cannot be easily distinguished on statistical
grounds alone. Although the \suzaku\ data below 1\,keV seems to favor
the relativistic reflection scenario, this energy range was excluded
from the fit due to concerns regarding the quality of the instrumental
calibration.

Since no model can be preferred based on the fit statistics, we have
discussed in detail the physical implications of these two models. 
In particular, we find that the quantities required to fit
the soft excess with the warm corona model---i.e., low temperature
($kT\sim0.5-1$\,keV) and large optical depth ($\tau_\mathrm{T}\sim10-20$)---
are incompatible with the physical concept of a corona, in which electron
scattering is expected to be the dominant source of opacity. Using
simple estimates of density, flux, and ionization
parameter, we have carried out calculations of plasmas in coronal and
photoionization equilibrium. In both cases, we found that atomic opacities
will dominate over Thomson opacities, predicting very strong absorption
features in the observed spectrum. Taking these simulations to the
most extreme cases, we find that it is very unlikely that a warm corona
can produce the soft featureless emission required to fit the data.

On the other hand, the relativistic reflection model appears more 
reasonable on physical grounds. Signatures of X-ray reflection have
been shown to be almost ubiquitous in most Seyfert AGN spectra, and
thus it is also expected to be present in \mrk509. The relativistic
reflection model requires extreme values for the spin and coronal
compactness, as well as a very large density for the reflector.
Although large densities are somewhat unexpected in accretion disks
around supermassive black holes, we cannot discard this possibility.
Therefore, based on the analysis presented here, we favor the 
high-density relativistic reflection scenario to explain the soft
excess in \mrk509.

Nonetheless, the present discussion is not entirely conclusive.  The
calculations described above do not include photon redistribution due to
Comptonization in the medium, nor any other source of turbulent motions capable
to broaden and smear the absorption lines present in the spectra. Evidently,
these effects are only relevant for the simulations at the highest temperatures
($T\sim10^7$\,K). For lower temperatures, the drastic modification of the
spectrum due to the strong absorption prevents this model to reproduce the soft
excess.  Detailed radiative transfer calculations covering larger optical
depths, Comptonization, velocity components, and the effects of the response of
current instruments are necessary to fully explore this problem. Such
calculations are well outside the scope of the present work, and thus will be
featured in a future publication. 

Finally, deeper observations of this source should be able to confirm or not
the presence of relativistic reflection. To clearly distinguish between the
narrow and broad components, future missions flying microcalorimeters such as
{\it XRISM} \citep{tas18}, {\it Athena} \citep{nan13}, and {\it Lynx}
\citep{oze18}, will become crucial.  However, in order to detect the shift of
Compton hump between the relativistic and the non-relativistic reflection,
focusing of hard photons with larger effective area than \nustar\ is necessary.
The concept mission {\it HEX-P} \citep{mad18}, will offer these capabilities.
Likewise, observations with instruments with broad-band coverage and good
sensitivity to both low and energies such as {\it STROBE-X} \citep{ray18} will
help to break model degeneracies further understand the nature of the soft
excess in \mrk509\ and many other AGN.

%
%
%
\acknowledgments 

We thank P.O.~Petrucci, J.~Malzac, B.~Czerny, A.~R{\'o}{\.z}a{\'n}ska, C.~Done,
and the members of the FERO collaboration for insightful discussions that
promoted many aspects of this paper. We also thank F.~Ursini for comments
to improve the manuscript.

J.A.G. acknowledges support from NASA grant NNX15AV31G and
from the Alexander von Humboldt Foundation.  R.M.T.C. has been supported by
NASA grant 80NSSC177K0515.  
E.G. acknowledges support by the DFG cluster of excellence ``Origin and Structure of the Universe".
M.B. acknowledges support from the Black Hole Initiative at Harvard
University, which is funded by a grant from the John Templeton Foundation.
J.F.S. has been supported by NASA Einstein
Fellowship grant PF5-160144.  F.T. acknowledges support by the Programma per
Giovani Ricercatori - anno 2014 ``Rita Levi Montalcini".
L.L. acknowledge support from NASA through grant number NNX15AP24G
C.R. acknowledges support from the CONICYT+PAI Convocatoria Nacional subvenci\'on
a instalaci\'on en la academia convocatoria a\~{n}o 2017 PAI77170080.

This work was partially supported under NASA contract No. NNG08FD60C and made use of data
from the {\it NuSTAR} mission, a project led by the California Institute of
Technology, managed by the Jet Propulsion Laboratory, and funded by the
National Aeronautics and Space Administration. We thank the {\it NuSTAR} Operations,
Software, and Calibration teams for support with the execution and analysis of
these observations. This research has made use of the {\it NuSTAR} Data Analysis
Software (NuSTARDAS), jointly developed by the ASI Science Data Center (ASDC,
Italy) and the California Institute of Technology (USA).

\vspace{5mm}
\facilities{\nustar, \suzaku\ (XIS)}
\software{{\sc xspec} \citep[v12.9.0d;][]{arn96}, 
{\sc MYtorus} \citep{mur09},
{\sc Borus02} \citep{bal18}, {\sc xillver} \citep{gar10,gar13a}, {\sc relxill}
\citep[v1.2.0;][]{gar14a,dau14}, {\sc xstar} \citep[v2.41;][]{kal01}, {\sc nustradas} (v1.6.0)}

%
%
%
%
\bibliographystyle{aasjournal}
\bibliography{my-references}

\begin{thebibliography}{}
\expandafter\ifx\csname natexlab\endcsname\relax\def\natexlab#1{#1}\fi
\providecommand{\url}[1]{\href{#1}{#1}}
\providecommand{\dodoi}[1]{doi:~\href{http://doi.org/#1}{\nolinkurl{#1}}}
\providecommand{\doeprint}[1]{\href{http://ascl.net/#1}{\nolinkurl{http://ascl.net/#1}}}
\providecommand{\doarXiv}[1]{\href{https://arxiv.org/abs/#1}{\nolinkurl{https://arxiv.org/abs/#1}}}

\bibitem[{{Arnaud}(1996)}]{arn96}
{Arnaud}, K.~A. 1996, in Astronomical Society of the Pacific Conference Series,
  Vol. 101, Astronomical Data Analysis Software and Systems V, ed. G.~H.
  {Jacoby} \& J.~{Barnes}, 17

\bibitem[{{Arnaud} {et~al.}(1985){Arnaud}, {Branduardi-Raymont}, {Culhane},
  {Fabian}, {Hazard}, {McGlynn}, {Shafer}, {Tennant}, \& {Ward}}]{arn85}
{Arnaud}, K.~A., {Branduardi-Raymont}, G., {Culhane}, J.~L., {et~al.} 1985,
  \mnras, 217, 105, \dodoi{10.1093/mnras/217.1.105}

\bibitem[{{Balokovi{\'c}} {et~al.}(2015){Balokovi{\'c}}, {Matt}, {Harrison},
  {Zoghbi}, {Ballantyne}, {Boggs}, {Christensen}, {Craig}, {Esmerian},
  {Fabian}, {F{\"u}rst}, {Hailey}, {Marinucci}, {Parker}, {Reynolds}, {Stern},
  {Walton}, \& {Zhang}}]{bal15}
{Balokovi{\'c}}, M., {Matt}, G., {Harrison}, F.~A., {et~al.} 2015, \apj, 800,
  62, \dodoi{10.1088/0004-637X/800/1/62}

\bibitem[{{Balokovi{\'c}} {et~al.}(2018){Balokovi{\'c}}, {Brightman},
  {Harrison}, {Comastri}, {Ricci}, {Buchner}, {Gandhi}, {Farrah}, \&
  {Stern}}]{bal18}
{Balokovi{\'c}}, M., {Brightman}, M., {Harrison}, F.~A., {et~al.} 2018, \apj,
  854, 42, \dodoi{10.3847/1538-4357/aaa7eb}

\bibitem[{{Boissay} {et~al.}(2016){Boissay}, {Ricci}, \& {Paltani}}]{boi16}
{Boissay}, R., {Ricci}, C., \& {Paltani}, S. 2016, \aap, 588, A70,
  \dodoi{10.1051/0004-6361/201526982}

\bibitem[{{Boissay} {et~al.}(2014){Boissay}, {Paltani}, {Ponti}, {Bianchi},
  {Cappi}, {Kaastra}, {Petrucci}, {Arav}, {Branduardi-Raymont}, {Costantini},
  {Ebrero}, {Kriss}, {Mehdipour}, {Pinto}, \& {Steenbrugge}}]{boi14}
{Boissay}, R., {Paltani}, S., {Ponti}, G., {et~al.} 2014, \aap, 567, A44,
  \dodoi{10.1051/0004-6361/201423494}

\bibitem[{{Brenneman} {et~al.}(2014){Brenneman}, {Madejski}, {Fuerst}, {Matt},
  {Elvis}, {Harrison}, {Ballantyne}, {Boggs}, {Christensen}, {Craig}, {Fabian},
  {Grefenstette}, {Hailey}, {Madsen}, {Marinucci}, {Rivers}, {Stern}, {Walton},
  \& {Zhang}}]{bre14}
{Brenneman}, L.~W., {Madejski}, G., {Fuerst}, F., {et~al.} 2014, \apj, 788, 61,
  \dodoi{10.1088/0004-637X/788/1/61}

\bibitem[{{Brightman} \& {Nandra}(2011)}]{bri11}
{Brightman}, M., \& {Nandra}, K. 2011, \mnras, 413, 1206,
  \dodoi{10.1111/j.1365-2966.2011.18207.x}

\bibitem[{{Cappi} {et~al.}(2009){Cappi}, {Tombesi}, {Bianchi}, {Dadina},
  {Giustini}, {Malaguti}, {Maraschi}, {Palumbo}, {Petrucci}, {Ponti},
  {Vignali}, \& {Yaqoob}}]{cap09}
{Cappi}, M., {Tombesi}, F., {Bianchi}, S., {et~al.} 2009, \aap, 504, 401,
  \dodoi{10.1051/0004-6361/200912137}

\bibitem[{{Costantini} {et~al.}(2016){Costantini}, {Kriss}, {Kaastra},
  {Bianchi}, {Branduardi-Raymont}, {Cappi}, {De Marco}, {Ebrero}, {Mehdipour},
  {Petrucci}, {Paltani}, {Ponti}, {Steenbrugge}, \& {Arav}}]{cos16}
{Costantini}, E., {Kriss}, G., {Kaastra}, J.~S., {et~al.} 2016, \aap, 595,
  A106, \dodoi{10.1051/0004-6361/201527956}

\bibitem[{{Crummy} {et~al.}(2006){Crummy}, {Fabian}, {Gallo}, \&
  {Ross}}]{cru06}
{Crummy}, J., {Fabian}, A.~C., {Gallo}, L., \& {Ross}, R.~R. 2006, \mnras, 365,
  1067, \dodoi{10.1111/j.1365-2966.2005.09844.x}

\bibitem[{{Czerny} \& {Elvis}(1987)}]{cze87}
{Czerny}, B., \& {Elvis}, M. 1987, \apj, 321, 305, \dodoi{10.1086/165630}

\bibitem[{{Czerny} {et~al.}(2003){Czerny}, {Niko{\l}ajuk},
  {R{\'o}{\.z}a{\'n}ska}, {Dumont}, {Loska}, \& {Zycki}}]{cze03}
{Czerny}, B., {Niko{\l}ajuk}, M., {R{\'o}{\.z}a{\'n}ska}, A., {et~al.} 2003,
  \aap, 412, 317, \dodoi{10.1051/0004-6361:20031441}

\bibitem[{{Dauser} {et~al.}(2014){Dauser}, {Garc\'ia}, {Parker}, {Fabian}, \&
  {Wimls}}]{dau14}
{Dauser}, T., {Garc\'ia}, J., {Parker}, M., {Fabian}, A., \& {Wimls}, J. 2014,
  Submitted to \mnras, 430, 1694

\bibitem[{{Dauser} {et~al.}(2016){Dauser}, {Garc{\'{\i}}a}, {Walton},
  {Eikmann}, {Kallman}, {McClintock}, \& {Wilms}}]{dau16}
{Dauser}, T., {Garc{\'{\i}}a}, J., {Walton}, D.~J., {et~al.} 2016, \aap, 590,
  A76, \dodoi{10.1051/0004-6361/201628135}

\bibitem[{{Dauser} {et~al.}(2013){Dauser}, {Garcia}, {Wilms}, {B{\"o}ck},
  {Brenneman}, {Falanga}, {Fukumura}, \& {Reynolds}}]{dau13}
{Dauser}, T., {Garcia}, J., {Wilms}, J., {et~al.} 2013, \mnras, 430, 1694,
  \dodoi{10.1093/mnras/sts710}

\bibitem[{{Done} {et~al.}(2012){Done}, {Davis}, {Jin}, {Blaes}, \&
  {Ward}}]{don12}
{Done}, C., {Davis}, S.~W., {Jin}, C., {Blaes}, O., \& {Ward}, M. 2012, \mnras,
  420, 1848, \dodoi{10.1111/j.1365-2966.2011.19779.x}

\bibitem[{{Fabian} {et~al.}(2002){Fabian}, {Ballantyne}, {Merloni}, {Vaughan},
  {Iwasawa}, \& {Boller}}]{fab02}
{Fabian}, A.~C., {Ballantyne}, D.~R., {Merloni}, A., {et~al.} 2002, \mnras,
  331, L35, \dodoi{10.1046/j.1365-8711.2002.05419.x}

\bibitem[{{Fabian} {et~al.}(2009){Fabian}, {Zoghbi}, {Ross}, {Uttley}, {Gallo},
  {Brandt}, {Blustin}, {Boller}, {Caballero-Garcia}, {Larsson}, {Miller},
  {Miniutti}, {Ponti}, {Reis}, {Reynolds}, {Tanaka}, \& {Young}}]{fab09}
{Fabian}, A.~C., {Zoghbi}, A., {Ross}, R.~R., {et~al.} 2009, \nat, 459, 540,
  \dodoi{10.1038/nature08007}

\bibitem[{{Fisher} {et~al.}(1995){Fisher}, {Huchra}, {Strauss}, {Davis},
  {Yahil}, \& {Schlegel}}]{fis95}
{Fisher}, K.~B., {Huchra}, J.~P., {Strauss}, M.~A., {et~al.} 1995, \apjs, 100,
  69, \dodoi{10.1086/192208}

\bibitem[{{Garc{\'{\i}}a} {et~al.}(2013){Garc{\'{\i}}a}, {Dauser}, {Reynolds},
  {Kallman}, {McClintock}, {Wilms}, \& {Eikmann}}]{gar13a}
{Garc{\'{\i}}a}, J., {Dauser}, T., {Reynolds}, C.~S., {et~al.} 2013, \apj, 768,
  146, \dodoi{10.1088/0004-637X/768/2/146}

\bibitem[{{Garc{\'{\i}}a} \& {Kallman}(2010)}]{gar10}
{Garc{\'{\i}}a}, J., \& {Kallman}, T.~R. 2010, \apj, 718, 695,
  \dodoi{10.1088/0004-637X/718/2/695}

\bibitem[{{Garc{\'{\i}}a} {et~al.}(2014){Garc{\'{\i}}a}, {Dauser}, {Lohfink},
  {Kallman}, {Steiner}, {McClintock}, {Brenneman}, {Wilms}, {Eikmann},
  {Reynolds}, \& {Tombesi}}]{gar14a}
{Garc{\'{\i}}a}, J., {Dauser}, T., {Lohfink}, A., {et~al.} 2014, \apj, 782, 76,
  \dodoi{10.1088/0004-637X/782/2/76}

\bibitem[{{Garc{\'{\i}}a} {et~al.}(2015){Garc{\'{\i}}a}, {Dauser}, {Steiner},
  {McClintock}, {Keck}, \& {Wilms}}]{gar15b}
{Garc{\'{\i}}a}, J.~A., {Dauser}, T., {Steiner}, J.~F., {et~al.} 2015, \apjl,
  808, L37, \dodoi{10.1088/2041-8205/808/2/L37}

\bibitem[{{Garc{\'{\i}}a} {et~al.}(2016){Garc{\'{\i}}a}, {Fabian}, {Kallman},
  {Dauser}, {Parker}, {McClintock}, {Steiner}, \& {Wilms}}]{gar16b}
{Garc{\'{\i}}a}, J.~A., {Fabian}, A.~C., {Kallman}, T.~R., {et~al.} 2016,
  \mnras, 462, 751, \dodoi{10.1093/mnras/stw1696}

\bibitem[{{Garc{\'\i}a} {et~al.}(2018){Garc{\'\i}a}, {Kallman}, {Bautista},
  {Mendoza}, {Deprince}, {Palmeri}, \& {Quinet}}]{gar18}
{Garc{\'\i}a}, J.~A., {Kallman}, T.~R., {Bautista}, M., {et~al.} 2018, in
  Astronomical Society of the Pacific Conference Series, Vol. 515, 282

\bibitem[{{George} \& {Fabian}(1991)}]{geo91}
{George}, I.~M., \& {Fabian}, A.~C. 1991, \mnras, 249, 352

\bibitem[{{Gierli{\'n}ski} \& {Done}(2004)}]{gie04b}
{Gierli{\'n}ski}, M., \& {Done}, C. 2004, \mnras, 349, L7,
  \dodoi{10.1111/j.1365-2966.2004.07687.x}

\bibitem[{{Haardt}(1993)}]{haa93}
{Haardt}, F. 1993, \apj, 413, 680, \dodoi{10.1086/173036}

\bibitem[{{Harrison} {et~al.}(2013){Harrison}, {Craig}, {Christensen},
  {Hailey}, {Zhang}, {Boggs}, {Stern}, {Cook}, {Forster}, {Giommi},
  {Grefenstette}, {Kim}, {Kitaguchi}, {Koglin}, {Madsen}, {Mao}, {Miyasaka},
  {Mori}, {Perri}, {Pivovaroff}, {Puccetti}, {Rana}, {Westergaard}, {Willis},
  {Zoglauer}, {An}, {Bachetti}, {Barri{\`e}re}, {Bellm}, {Bhalerao},
  {Brejnholt}, {Fuerst}, {Liebe}, {Markwardt}, {Nynka}, {Vogel}, {Walton},
  {Wik}, {Alexander}, {Cominsky}, {Hornschemeier}, {Hornstrup}, {Kaspi},
  {Madejski}, {Matt}, {Molendi}, {Smith}, {Tomsick}, {Ajello}, {Ballantyne},
  {Balokovi{\'c}}, {Barret}, {Bauer}, {Blandford}, {Brandt}, {Brenneman},
  {Chiang}, {Chakrabarty}, {Chenevez}, {Comastri}, {Dufour}, {Elvis}, {Fabian},
  {Farrah}, {Fryer}, {Gotthelf}, {Grindlay}, {Helfand}, {Krivonos}, {Meier},
  {Miller}, {Natalucci}, {Ogle}, {Ofek}, {Ptak}, {Reynolds}, {Rigby},
  {Tagliaferri}, {Thorsett}, {Treister}, \& {Urry}}]{har13}
{Harrison}, F.~A., {Craig}, W.~W., {Christensen}, F.~E., {et~al.} 2013, \apj,
  770, 103, \dodoi{10.1088/0004-637X/770/2/103}

\bibitem[{{Jiang} {et~al.}(2018){Jiang}, {Parker}, {Fabian}, {Alston},
  {Buisson}, {Cackett}, {Chiang}, {Dauser}, {Gallo}, {Garc{\'{\i}}a},
  {Harrison}, {Lohfink}, {De Marco}, {Kara}, {Miller}, {Miniutti}, {Pinto},
  {Walton}, \& {Wilkins}}]{jia18}
{Jiang}, J., {Parker}, M.~L., {Fabian}, A.~C., {et~al.} 2018, \mnras, 477,
  3711, \dodoi{10.1093/mnras/sty836}

\bibitem[{{Jin} {et~al.}(2009){Jin}, {Done}, {Ward}, {Gierli{\'n}ski}, \&
  {Mullaney}}]{jin09}
{Jin}, C., {Done}, C., {Ward}, M., {Gierli{\'n}ski}, M., \& {Mullaney}, J.
  2009, \mnras, 398, L16, \dodoi{10.1111/j.1745-3933.2009.00697.x}

\bibitem[{{Kaastra} {et~al.}(2011){Kaastra}, {Petrucci}, {Cappi}, {Arav},
  {Behar}, {Bianchi}, {Bloom}, {Blustin}, {Branduardi-Raymont}, {Costantini},
  {Dadina}, {Detmers}, {Ebrero}, {Jonker}, {Klein}, {Kriss}, {Lubi{\'n}ski},
  {Malzac}, {Mehdipour}, {Paltani}, {Pinto}, {Ponti}, {Ratti}, {Smith},
  {Steenbrugge}, \& {de Vries}}]{kaa11}
{Kaastra}, J.~S., {Petrucci}, P.-O., {Cappi}, M., {et~al.} 2011, \aap, 534,
  A36, \dodoi{10.1051/0004-6361/201116869}

\bibitem[{{Kaastra} {et~al.}(2014){Kaastra}, {Ebrero}, {Arav}, {Behar},
  {Bianchi}, {Branduardi-Raymont}, {Cappi}, {Costantini}, {Kriss}, {De Marco},
  {Mehdipour}, {Paltani}, {Petrucci}, {Pinto}, {Ponti}, {Steenbrugge}, \& {de
  Vries}}]{kaa14}
{Kaastra}, J.~S., {Ebrero}, J., {Arav}, N., {et~al.} 2014, \aap, 570, A73,
  \dodoi{10.1051/0004-6361/201424662}

\bibitem[{{Kalberla} {et~al.}(2005){Kalberla}, {Burton}, {Hartmann}, {Arnal},
  {Bajaja}, {Morras}, \& {P{\"o}ppel}}]{kal05}
{Kalberla}, P.~M.~W., {Burton}, W.~B., {Hartmann}, D., {et~al.} 2005, \aap,
  440, 775, \dodoi{10.1051/0004-6361:20041864}

\bibitem[{{Kallman} \& {Bautista}(2001)}]{kal01}
{Kallman}, T., \& {Bautista}, M. 2001, \apjs, 133, 221, \dodoi{10.1086/319184}

\bibitem[{{Kara} {et~al.}(2017){Kara}, {Garc{\'{\i}}a}, {Lohfink}, {Fabian},
  {Reynolds}, {Tombesi}, \& {Wilkins}}]{kar17}
{Kara}, E., {Garc{\'{\i}}a}, J.~A., {Lohfink}, A., {et~al.} 2017, \mnras, 468,
  3489, \dodoi{10.1093/mnras/stx792}

\bibitem[{{Kaufman} {et~al.}(2018){Kaufman}, {Blaes}, \& {Hirose}}]{kau18}
{Kaufman}, J., {Blaes}, O.~M., \& {Hirose}, S. 2018, \mnras, 476, 5548,
  \dodoi{10.1093/mnras/sty540}

\bibitem[{{Keck} {et~al.}(2014){Keck}, {Brenneman}, {Ballantyne}, {Bauer},
  {Boggs}, {Christensen}, {Craig}, {Dauser}, {Elvis}, {Fabian}, {Fuerst},
  {Garc\'ia}, \& {et al.}}]{kec14}
{Keck}, M., {Brenneman}, L., {Ballantyne}, D., {et~al.} 2014

\bibitem[{{Keek} \& {Ballantyne}(2016)}]{kee16}
{Keek}, L., \& {Ballantyne}, D.~R. 2016, \mnras, 456, 2722,
  \dodoi{10.1093/mnras/stv2882}

\bibitem[{{Kettula} {et~al.}(2013){Kettula}, {Nevalainen}, \& {Miller}}]{ket13}
{Kettula}, K., {Nevalainen}, J., \& {Miller}, E.~D. 2013, \aap, 552, A47,
  \dodoi{10.1051/0004-6361/201220408}

\bibitem[{{Koyama} {et~al.}(2007){Koyama}, {Tsunemi}, {Dotani}, {Bautz},
  {Hayashida}, {Tsuru}, {Matsumoto}, {Ogawara}, {Ricker}, {Doty}, {Kissel},
  {Foster}, {Nakajima}, {Yamaguchi}, {Mori}, {Sakano}, {Hamaguchi},
  {Nishiuchi}, {Miyata}, {Torii}, {Namiki}, {Katsuda}, {Matsuura}, {Miyauchi},
  {Anabuki}, {Tawa}, {Ozaki}, {Murakami}, {Maeda}, {Ichikawa}, {Prigozhin},
  {Boughan}, {Lamarr}, {Miller}, {Burke}, {Gregory}, {Pillsbury}, {Bamba},
  {Hiraga}, {Senda}, {Katayama}, {Kitamoto}, {Tsujimoto}, {Kohmura}, {Tsuboi},
  \& {Awaki}}]{koy07}
{Koyama}, K., {Tsunemi}, H., {Dotani}, T., {et~al.} 2007, \pasj, 59, 23,
  \dodoi{10.1093/pasj/59.sp1.S23}

\bibitem[{{Krolik} \& {Kallman}(1984)}]{kro84}
{Krolik}, J.~H., \& {Kallman}, T.~R. 1984, \apj, 286, 366,
  \dodoi{10.1086/162608}

\bibitem[{{Laor}(1991)}]{lao91}
{Laor}, A. 1991, \apj, 376, 90, \dodoi{10.1086/170257}

\bibitem[{{Leighly}(1999)}]{lei99}
{Leighly}, K.~M. 1999, \apjs, 125, 317, \dodoi{10.1086/313287}

\bibitem[{{Liu} \& {Li}(2015)}]{liu15}
{Liu}, Y., \& {Li}, X. 2015, \mnras, 448, L53, \dodoi{10.1093/mnrasl/slu198}

\bibitem[{{Madsen} {et~al.}(2017){Madsen}, {Beardmore}, {Forster}, {Guainazzi},
  {Marshall}, {Miller}, {Page}, \& {Stuhlinger}}]{mad17}
{Madsen}, K.~K., {Beardmore}, A.~P., {Forster}, K., {et~al.} 2017, \aj, 153, 2,
  \dodoi{10.3847/1538-3881/153/1/2}

\bibitem[{{Madsen} {et~al.}(2015){Madsen}, {Harrison}, {Markwardt}, {An},
  {Grefenstette}, {Bachetti}, {Miyasaka}, {Kitaguchi}, {Bhalerao}, {Boggs},
  {Christensen}, {Craig}, {Forster}, {Fuerst}, {Hailey}, {Perri}, {Puccetti},
  {Rana}, {Stern}, {Walton}, {J{\o}rgen Westergaard}, \& {Zhang}}]{mad15}
{Madsen}, K.~K., {Harrison}, F.~A., {Markwardt}, C.~B., {et~al.} 2015, \apjs,
  220, 8, \dodoi{10.1088/0067-0049/220/1/8}

\bibitem[{{Madsen} {et~al.}(2018){Madsen}, {Harrison}, {Broadway},
  {Christensen}, {Descalle}, {Ferreira}, {Grefenstette}, {Gurgew},
  {Hornschemeier}, {Miyasaka}, {Okajima}, {Pike}, {Pivovaroff}, {Saha},
  {Stern}, {Vogel}, {Windt}, \& {Zhang}}]{mad18}
{Madsen}, K.~K., {Harrison}, F., {Broadway}, D., {et~al.} 2018, in Society of
  Photo-Optical Instrumentation Engineers (SPIE) Conference Series, Vol. 10699,
  Society of Photo-Optical Instrumentation Engineers (SPIE) Conference Series,
  106996M

\bibitem[{{Magdziarz} {et~al.}(1998){Magdziarz}, {Blaes}, {Zdziarski},
  {Johnson}, \& {Smith}}]{mag98}
{Magdziarz}, P., {Blaes}, O.~M., {Zdziarski}, A.~A., {Johnson}, W.~N., \&
  {Smith}, D.~A. 1998, \mnras, 301, 179,
  \dodoi{10.1046/j.1365-8711.1998.02015.x}

\bibitem[{{Mallick} {et~al.}(2018){Mallick}, {Alston}, {Parker}, {Fabian},
  {Pinto}, {Dewangan}, {Markowitz}, {Gandhi}, {Kembhavi}, \& {Misra}}]{mal18}
{Mallick}, L., {Alston}, W.~N., {Parker}, M.~L., {et~al.} 2018, \mnras, 479,
  615, \dodoi{10.1093/mnras/sty1487}

\bibitem[{{Marinucci} {et~al.}(2018){Marinucci}, {Bianchi}, {Braito}, {Matt},
  {Nardini}, \& {Reeves}}]{mar18}
{Marinucci}, A., {Bianchi}, S., {Braito}, V., {et~al.} 2018, \mnras, 478, 5638,
  \dodoi{10.1093/mnras/sty1436}

\bibitem[{{Marinucci} {et~al.}(2016){Marinucci}, {Tortosa}, \& {NuSTAR AGN
  Physics Working Group}}]{mar16}
{Marinucci}, A., {Tortosa}, A., \& {NuSTAR AGN Physics Working Group}. 2016,
  Astronomische Nachrichten, 337, 490, \dodoi{10.1002/asna.201612335}

\bibitem[{{Markoff} {et~al.}(2005){Markoff}, {Nowak}, \& {Wilms}}]{mar05}
{Markoff}, S., {Nowak}, M.~A., \& {Wilms}, J. 2005, 635, 1203

\bibitem[{{Matt} {et~al.}(1991){Matt}, {Perola}, \& {Piro}}]{mat91}
{Matt}, G., {Perola}, G.~C., \& {Piro}, L. 1991, \aap, 247, 25

\bibitem[{{Matt} {et~al.}(1992){Matt}, {Perola}, {Piro}, \& {Stella}}]{mat92}
{Matt}, G., {Perola}, G.~C., {Piro}, L., \& {Stella}, L. 1992, \aap, 257, 63

\bibitem[{{Mehdipour} {et~al.}(2011){Mehdipour}, {Branduardi-Raymont},
  {Kaastra}, {Petrucci}, {Kriss}, {Ponti}, {Blustin}, {Paltani}, {Cappi},
  {Detmers}, \& {Steenbrugge}}]{meh11}
{Mehdipour}, M., {Branduardi-Raymont}, G., {Kaastra}, J.~S., {et~al.} 2011,
  \aap, 534, A39, \dodoi{10.1051/0004-6361/201116875}

\bibitem[{{Middleton} {et~al.}(2009){Middleton}, {Done}, {Ward},
  {Gierli{\'n}ski}, \& {Schurch}}]{mid09}
{Middleton}, M., {Done}, C., {Ward}, M., {Gierli{\'n}ski}, M., \& {Schurch}, N.
  2009, \mnras, 394, 250, \dodoi{10.1111/j.1365-2966.2008.14255.x}

\bibitem[{{Miniutti} {et~al.}(2009){Miniutti}, {Ponti}, {Greene}, {Ho},
  {Fabian}, \& {Iwasawa}}]{min09}
{Miniutti}, G., {Ponti}, G., {Greene}, J.~E., {et~al.} 2009, \mnras, 394, 443,
  \dodoi{10.1111/j.1365-2966.2008.14334.x}

\bibitem[{{Morini} {et~al.}(1987){Morini}, {Lipani}, \& {Molteni}}]{mor87}
{Morini}, M., {Lipani}, N.~A., \& {Molteni}, D. 1987, \apj, 317, 145,
  \dodoi{10.1086/165262}

\bibitem[{{Murphy} \& {Yaqoob}(2009)}]{mur09}
{Murphy}, K.~D., \& {Yaqoob}, T. 2009, \mnras, 397, 1549,
  \dodoi{10.1111/j.1365-2966.2009.15025.x}

\bibitem[{{Nandra} {et~al.}(2013){Nandra}, {Barret}, {Barcons}, {Fabian}, {den
  Herder}, {Piro}, {Watson}, {Adami}, {Aird}, {Afonso}, {Alexander},
  {Argiroffi}, {Amati}, {Arnaud}, {Atteia}, {Audard}, {Badenes}, {Ballet},
  {Ballo}, {Bamba}, {Bhardwaj}, {Stefano Battistelli}, {Becker}, {De Becker},
  {Behar}, {Bianchi}, {Biffi}, {B{\^\i}rzan}, {Bocchino}, {Bogdanov}, {Boirin},
  {Boller}, {Borgani}, {Borm}, {Bouch{\'e}}, {Bourdin}, {Bower}, {Braito},
  {Branchini}, {Branduardi-Raymont}, {Bregman}, {Brenneman}, {Brightman},
  {Br{\"u}ggen}, {Buchner}, {Bulbul}, {Brusa}, {Bursa}, {Caccianiga},
  {Cackett}, {Campana}, {Cappelluti}, {Cappi}, {Carrera}, {Ceballos},
  {Christensen}, {Chu}, {Churazov}, {Clerc}, {Corbel}, {Corral}, {Comastri},
  {Costantini}, {Croston}, {Dadina}, {D'Ai}, {Decourchelle}, {Della Ceca},
  {Dennerl}, {Dolag}, {Done}, {Dovciak}, {Drake}, {Eckert}, {Edge}, {Ettori},
  {Ezoe}, {Feigelson}, {Fender}, {Feruglio}, {Finoguenov}, {Fiore}, {Galeazzi},
  {Gallagher}, {Gandhi}, {Gaspari}, {Gastaldello}, {Georgakakis},
  {Georgantopoulos}, {Gilfanov}, {Gitti}, {Gladstone}, {Goosmann}, {Gosset},
  {Grosso}, {Guedel}, {Guerrero}, {Haberl}, {Hardcastle}, {Heinz}, {Alonso
  Herrero}, {Herv{\'e}}, {Holmstrom}, {Iwasawa}, {Jonker}, {Kaastra}, {Kara},
  {Karas}, {Kastner}, {King}, {Kosenko}, {Koutroumpa}, {Kraft}, {Kreykenbohm},
  {Lallement}, {Lanzuisi}, {Lee}, {Lemoine- Goumard}, {Lobban}, {Lodato},
  {Lovisari}, {Lotti}, {McCharthy}, {McNamara}, {Maggio}, {Maiolino}, {De
  Marco}, {de Martino}, {Mateos}, {Matt}, {Maughan}, {Mazzotta}, {Mendez},
  {Merloni}, {Micela}, {Miceli}, {Mignani}, {Miller}, {Miniutti}, {Molendi},
  {Montez}, {Moretti}, {Motch}, {Naz{\'e}}, {Nevalainen}, {Nicastro}, {Nulsen},
  {Ohashi}, {O'Brien}, {Osborne}, {Oskinova}, {Pacaud}, {Paerels}, {Page},
  {Papadakis}, {Pareschi}, {Petre}, {Petrucci}, {Piconcelli}, {Pillitteri},
  {Pinto}, {de Plaa}, {Pointecouteau}, {Ponman}, {Ponti}, {Porquet}, {Pounds},
  {Pratt}, {Predehl}, {Proga}, {Psaltis}, {Rafferty}, {Ramos-Ceja}, {Ranalli},
  {Rasia}, {Rau}, {Rauw}, {Rea}, {Read}, {Reeves}, {Reiprich}, {Renaud},
  {Reynolds}, {Risaliti}, {Rodriguez}, {Rodriguez Hidalgo}, {Roncarelli},
  {Rosario}, {Rossetti}, {Rozanska}, {Rovilos}, {Salvaterra}, {Salvato}, {Di
  Salvo}, {Sanders}, {Sanz-Forcada}, {Schawinski}, {Schaye}, {Schwope},
  {Sciortino}, {Severgnini}, {Shankar}, {Sijacki}, {Sim}, {Schmid}, {Smith},
  {Steiner}, {Stelzer}, {Stewart}, {Strohmayer}, {Str{\"u}der}, {Sun}, {Takei},
  {Tatischeff}, {Tiengo}, {Tombesi}, {Trinchieri}, {Tsuru}, {Ud-Doula},
  {Ursino}, {Valencic}, {Vanzella}, {Vaughan}, {Vignali}, {Vink}, {Vito},
  {Volonteri}, {Wang}, {Webb}, {Willingale}, {Wilms}, {Wise}, {Worrall},
  {Young}, {Zampieri}, {In't Zand}, {Zane}, {Zezas}, {Zhang}, \&
  {Zhuravleva}}]{nan13}
{Nandra}, K., {Barret}, D., {Barcons}, X., {et~al.} 2013, ArXiv e-prints,
  arXiv:1306.2307.
\newblock \doarXiv{1306.2307}

\bibitem[{{Nardini} {et~al.}(2012){Nardini}, {Fabian}, \& {Walton}}]{nard12}
{Nardini}, E., {Fabian}, A.~C., \& {Walton}, D.~J. 2012, \mnras, 423, 3299,
  \dodoi{10.1111/j.1365-2966.2012.21123.x}

\bibitem[{{Nardini} {et~al.}(2016){Nardini}, {Porquet}, {Reeves}, {Braito},
  {Lobban}, \& {Matt}}]{nar16}
{Nardini}, E., {Porquet}, D., {Reeves}, J.~N., {et~al.} 2016, \apj, 832, 45,
  \dodoi{10.3847/0004-637X/832/1/45}

\bibitem[{{{\"O}zel}(2018)}]{oze18}
{{\"O}zel}, F. 2018, Nature Astronomy, 2, 608,
  \dodoi{10.1038/s41550-018-0548-3}

\bibitem[{{Parker} {et~al.}(2018){Parker}, {Miller}, \& {Fabian}}]{par18}
{Parker}, M.~L., {Miller}, J.~M., \& {Fabian}, A.~C. 2018, \mnras, 474, 1538,
  \dodoi{10.1093/mnras/stx2861}

\bibitem[{{Parker} {et~al.}(2017){Parker}, {Pinto}, {Fabian}, {Lohfink},
  {Buisson}, {Alston}, {Kara}, {Cackett}, {Chiang}, {Dauser}, {De Marco},
  {Gallo}, {Garcia}, {Harrison}, {King}, {Middleton}, {Miller}, {Miniutti},
  {Reynolds}, {Uttley}, {Vasudevan}, {Walton}, {Wilkins}, \& {Zoghbi}}]{par17}
{Parker}, M.~L., {Pinto}, C., {Fabian}, A.~C., {et~al.} 2017, \nat, 543, 83,
  \dodoi{10.1038/nature21385}

\bibitem[{{Peterson} {et~al.}(2004){Peterson}, {Ferrarese}, {Gilbert}, {Kaspi},
  {Malkan}, {Maoz}, {Merritt}, {Netzer}, {Onken}, {Pogge}, {Vestergaard}, \&
  {Wandel}}]{pet04}
{Peterson}, B.~M., {Ferrarese}, L., {Gilbert}, K.~M., {et~al.} 2004, \apj, 613,
  682, \dodoi{10.1086/423269}

\bibitem[{{Petrucci} {et~al.}(2001){Petrucci}, {Merloni}, {Fabian}, {Haardt},
  \& {Gallo}}]{pet01}
{Petrucci}, P.~O., {Merloni}, A., {Fabian}, A., {Haardt}, F., \& {Gallo}, E.
  2001, \mnras, 328, 501, \dodoi{10.1046/j.1365-8711.2001.04897.x}

\bibitem[{{Petrucci} {et~al.}(2018){Petrucci}, {Ursini}, {De Rosa}, {Bianchi},
  {Cappi}, {Matt}, {Dadina}, \& {Malzac}}]{pet18}
{Petrucci}, P.-O., {Ursini}, F., {De Rosa}, A., {et~al.} 2018, \aap, 611, A59,
  \dodoi{10.1051/0004-6361/201731580}

\bibitem[{{Petrucci} {et~al.}(2013){Petrucci}, {Paltani}, {Malzac}, {Kaastra},
  {Cappi}, {Ponti}, {De Marco}, {Kriss}, {Steenbrugge}, {Bianchi},
  {Branduardi-Raymont}, {Mehdipour}, {Costantini}, {Dadina}, \&
  {Lubi{\'n}ski}}]{pet13}
{Petrucci}, P.-O., {Paltani}, S., {Malzac}, J., {et~al.} 2013, \aap, 549, A73,
  \dodoi{10.1051/0004-6361/201219956}

\bibitem[{{Piconcelli} {et~al.}(2005){Piconcelli}, {Jimenez-Bail{\'o}n},
  {Guainazzi}, {Schartel}, {Rodr{\'{\i}}guez-Pascual}, \&
  {Santos-Lle{\'o}}}]{pic05}
{Piconcelli}, E., {Jimenez-Bail{\'o}n}, E., {Guainazzi}, M., {et~al.} 2005,
  \aap, 432, 15, \dodoi{10.1051/0004-6361:20041621}

\bibitem[{{Ponti} {et~al.}(2013){Ponti}, {Cappi}, {Costantini}, {Bianchi},
  {Kaastra}, {De Marco}, {Fender}, {Petrucci}, {Kriss}, {Steenbrugge}, {Arav},
  {Behar}, {Branduardi-Raymont}, {Dadina}, {Ebrero}, {Lubi{\'n}ski},
  {Mehdipour}, {Paltani}, {Pinto}, \& {Tombesi}}]{pon13}
{Ponti}, G., {Cappi}, M., {Costantini}, E., {et~al.} 2013, \aap, 549, A72,
  \dodoi{10.1051/0004-6361/201219450}

\bibitem[{{Porquet} {et~al.}(2004){Porquet}, {Reeves}, {O'Brien}, \&
  {Brinkmann}}]{por04}
{Porquet}, D., {Reeves}, J.~N., {O'Brien}, P., \& {Brinkmann}, W. 2004, \aap,
  422, 85, \dodoi{10.1051/0004-6361:20047108}

\bibitem[{{Porquet} {et~al.}(2018){Porquet}, {Reeves}, {Matt}, {Marinucci},
  {Nardini}, {Braito}, {Lobban}, {Ballantyne}, {Boggs}, {Christensen},
  {Dauser}, {Farrah}, {Garcia}, {Hailey}, {Harrison}, {Stern}, {Tortosa},
  {Ursini}, \& {Zhang}}]{por18}
{Porquet}, D., {Reeves}, J.~N., {Matt}, G., {et~al.} 2018, \aap, 609, A42,
  \dodoi{10.1051/0004-6361/201731290}

\bibitem[{{Pounds} {et~al.}(1994){Pounds}, {Nandra}, {Fink}, \&
  {Makino}}]{pou94}
{Pounds}, K.~A., {Nandra}, K., {Fink}, H.~H., \& {Makino}, F. 1994, \mnras,
  267, 193

\bibitem[{{Pounds} {et~al.}(1986){Pounds}, {Warwick}, {Culhane}, \& {de
  Korte}}]{pou86}
{Pounds}, K.~A., {Warwick}, R.~S., {Culhane}, J.~L., \& {de Korte}, P.~A.~J.
  1986, \mnras, 218, 685, \dodoi{10.1093/mnras/218.4.685}

\bibitem[{{Ray} {et~al.}(2018){Ray}, {Arzoumanian}, {Brandt}, {Burns},
  {Chakrabarty}, {Feroci}, {Gendreau}, {Gevin}, {Hernanz}, {Jenke}, {Kenyon},
  {G{\'a}lvez}, {Maccarone}, {Okajima}, {Remillard}, {Schanne}, {Tenzer},
  {Vacchi}, {Wilson-Hodge}, {Winter}, {Zane}, {Ballantyne}, {Bozzo},
  {Brenneman}, {Cackett}, {De Rosa}, {Goldstein}, {Hartmann}, {McDonald},
  {Stevens}, {Tomsick}, {Watts}, {Wood}, \& {Zoghbi}}]{ray18}
{Ray}, P.~S., {Arzoumanian}, Z., {Brandt}, S., {et~al.} 2018, in Society of
  Photo-Optical Instrumentation Engineers (SPIE) Conference Series, Vol. 10699,
  Society of Photo-Optical Instrumentation Engineers (SPIE) Conference Series,
  1069919

\bibitem[{{Ricci} {et~al.}(2014){Ricci}, {Ueda}, {Paltani}, {Ichikawa},
  {Gandhi}, \& {Awaki}}]{ric14}
{Ricci}, C., {Ueda}, Y., {Paltani}, S., {et~al.} 2014, \mnras, 441, 3622,
  \dodoi{10.1093/mnras/stu735}

\bibitem[{{Ricci} {et~al.}(2017){Ricci}, {Trakhtenbrot}, {Koss}, {Ueda}, {Del
  Vecchio}, {Treister}, {Schawinski}, {Paltani}, {Oh}, {Lamperti}, {Berney},
  {Gandhi}, {Ichikawa}, {Bauer}, {Ho}, {Asmus}, {Beckmann}, {Soldi},
  {Balokovi{\'c}}, {Gehrels}, \& {Markwardt}}]{ric17}
{Ricci}, C., {Trakhtenbrot}, B., {Koss}, M.~J., {et~al.} 2017, \apjs, 233, 17,
  \dodoi{10.3847/1538-4365/aa96ad}

\bibitem[{{Risaliti} {et~al.}(2013){Risaliti}, {Harrison}, {Madsen}, {Walton},
  {Boggs}, {Christensen}, {Craig}, {Grefenstette}, {Hailey}, {Nardini},
  {Stern}, \& {Zhang}}]{ris13}
{Risaliti}, G., {Harrison}, F.~A., {Madsen}, K.~K., {et~al.} 2013, \nat, 494,
  449, \dodoi{10.1038/nature11938}

\bibitem[{{Ross} \& {Fabian}(2005)}]{ros05}
{Ross}, R.~R., \& {Fabian}, A.~C. 2005, \mnras, 358, 211,
  \dodoi{10.1111/j.1365-2966.2005.08797.x}

\bibitem[{{Ross} {et~al.}(1978){Ross}, {Weaver}, \& {McCray}}]{ros78}
{Ross}, R.~R., {Weaver}, R., \& {McCray}, R. 1978, \apj, 219, 292,
  \dodoi{10.1086/155776}

\bibitem[{{R{\'o}{\.z}a{\'n}ska} {et~al.}(2015){R{\'o}{\.z}a{\'n}ska},
  {Malzac}, {Belmont}, {Czerny}, \& {Petrucci}}]{roz15}
{R{\'o}{\.z}a{\'n}ska}, A., {Malzac}, J., {Belmont}, R., {Czerny}, B., \&
  {Petrucci}, P.-O. 2015, \aap, 580, A77, \dodoi{10.1051/0004-6361/201526288}

\bibitem[{{Shakura} \& {Sunyaev}(1973)}]{sha73}
{Shakura}, N.~I., \& {Sunyaev}, R.~A. 1973, \aa, 24, 337

\bibitem[{{Singh} {et~al.}(1985){Singh}, {Garmire}, \& {Nousek}}]{sin85}
{Singh}, K.~P., {Garmire}, G.~P., \& {Nousek}, J. 1985, \apj, 297, 633,
  \dodoi{10.1086/163560}

\bibitem[{{Smith} {et~al.}(2007){Smith}, {Page}, \&
  {Branduardi-Raymont}}]{smi07}
{Smith}, R.~A.~N., {Page}, M.~J., \& {Branduardi-Raymont}, G. 2007, \aap, 461,
  135, \dodoi{10.1051/0004-6361:20065348}

\bibitem[{{Svensson} \& {Zdziarski}(1994)}]{sve94}
{Svensson}, R., \& {Zdziarski}, A.~A. 1994, \apj, 436, 599,
  \dodoi{10.1086/174934}

\bibitem[{{Tashiro} {et~al.}(2018){Tashiro}, {Maejima}, {Toda}, {Kelley},
  {Reichenthal}, {Lobell}, {Petre}, {Guainazzi}, {Costantini}, {Edison},
  {Fujimoto}, {Grim}, {Hayashida}, {den Herder}, {Ishisaki}, {Paltani},
  {Matsushita}, {Mori}, {Sneiderman}, {Takei}, {Terada}, {Tomida}, {Akamatsu},
  {Angelini}, {Arai}, {Awaki}, {Babyk}, {Bamba}, {Barfknecht}, {Barnstable},
  {Bialas}, {Blagojevic}, {Bonafede}, {Brambora}, {Brenneman}, {Brown},
  {Brown}, {Burns}, {Canavan}, {Carnahan}, {Chiao}, {Comber}, {Corrales}, {de
  Vries}, {Dercksen}, {Diaz-Trigo}, {Dillard}, {DiPirro}, {Done}, {Dotani},
  {Ebisawa}, {Eckart}, {Enoto}, {Ezoe}, {Ferrigno}, {Fukazawa}, {Fujita},
  {Furuzawa}, {Gallo}, {Graham}, {Gu}, {Hagino}, {Hamaguchi}, {Hatsukade},
  {Hawes}, {Hayashi}, {Hegarty}, {Hell}, {Hiraga}, {Hodges-Kluck}, {Holland},
  {Hornschemeier}, {Hoshino}, {Ichinohe}, {Iizuka}, {Ishibashi}, {Ishida},
  {Ishikawa}, {Ishimura}, {James}, {Kallman}, {Kara}, {Katsuda}, {Kenyon},
  {Kilbourne}, {Kimball}, {Kitaguti}, {Kitamoto}, {Kobayashi}, {Kohmura},
  {Koyama}, {Kubota}, {Leutenegger}, {Lockard}, {Loewenstein}, {Maeda},
  {Marbley}, {Markevitch}, {Matsumoto}, {Matsuzaki}, {McCammon}, {McNamara},
  {Miko}, {Miller}, {Miller}, {Minesugi}, {Mitsuishi}, {Mizuno}, {Mori},
  {Mukai}, {Murakami}, {Mushotzky}, {Nakajima}, {Nakamura}, {Nakashima},
  {Nakazawa}, {Natsukari}, {Nigo}, {Nishioka}, {Nobukawa}, {Nobukawa}, {Noda},
  {Odaka}, {Ogawa}, {Ohashi}, {Ohno}, {Ohta}, {Okajima}, {Okamoto}, {Onizuka},
  {Ota}, {Ozaki}, {Plucinsky}, {Porter}, {Pottschmidt}, {Sato}, {Sato},
  {Sawada}, {Seta}, {Shelton}, {Shibano}, {Shida}, {Shidatsu}, {Shirron},
  {Simionescu}, {Smith}, {Someya}, {Soong}, {Suagawara}, {Szymkowiak},
  {Takahashi}, {Tamagawa}, {Tamura}, {Tanaka}, {Terashima}, {Tsuboi},
  {Tsujimoto}, {Tsunemi}, {Tsuru}, {Uchida}, {Uchiyama}, {Ueda}, {Uno},
  {Walsh}, {Watanabe}, {Williams}, {Wolfs}, {Wright}, {Yamada}, {Yamaguchi},
  {Yamaoka}, {Yamasaki}, {Yamauchi}, {Yamauchi}, {Yanagase}, {Yaqoob},
  {Yasuda}, {Yoshioka}, {Zabala}, \& {Irina}}]{tas18}
{Tashiro}, M., {Maejima}, H., {Toda}, K., {et~al.} 2018, in Society of
  Photo-Optical Instrumentation Engineers (SPIE) Conference Series, Vol. 10699,
  1069922

\bibitem[{{Tomsick} {et~al.}(2018){Tomsick}, {Parker}, {Garc{\'{\i}}a},
  {Yamaoka}, {Barret}, {Chiu}, {Clavel}, {Fabian}, {F{\"u}rst}, {Gandhi},
  {Grinberg}, {Miller}, {Pottschmidt}, \& {Walton}}]{tom18}
{Tomsick}, J.~A., {Parker}, M.~L., {Garc{\'{\i}}a}, J.~A., {et~al.} 2018, \apj,
  855, 3, \dodoi{10.3847/1538-4357/aaaab1}

\bibitem[{{Tortosa} {et~al.}(2018){Tortosa}, {Bianchi}, {Marinucci}, {Matt}, \&
  {Petrucci}}]{tor18}
{Tortosa}, A., {Bianchi}, S., {Marinucci}, A., {Matt}, G., \& {Petrucci}, P.~O.
  2018, \aap, 614, A37, \dodoi{10.1051/0004-6361/201732382}

\bibitem[{{Tortosa} {et~al.}(2017){Tortosa}, {Marinucci}, {Matt}, {Bianchi},
  {La Franca}, {Ballantyne}, {Boorman}, {Fabian}, {Farrah}, {Fuerst}, {Gandhi},
  {Harrison}, {Koss}, {Ricci}, {Stern}, {Ursini}, \& {Walton}}]{tor17}
{Tortosa}, A., {Marinucci}, A., {Matt}, G., {et~al.} 2017, \mnras, 466, 4193,
  \dodoi{10.1093/mnras/stw3301}

\bibitem[{{Ursini} {et~al.}(2015){Ursini}, {Boissay}, {Petrucci}, {Matt},
  {Cappi}, {Bianchi}, {Kaastra}, {Harrison}, {Walton}, {di Gesu}, {Costantini},
  {De Marco}, {Kriss}, {Mehdipour}, {Paltani}, {Peterson}, {Ponti}, \&
  {Steenbrugge}}]{urs15}
{Ursini}, F., {Boissay}, R., {Petrucci}, P.-O., {et~al.} 2015, \aap, 577, A38,
  \dodoi{10.1051/0004-6361/201425401}

\bibitem[{{Vasudevan} \& {Fabian}(2007)}]{vas07}
{Vasudevan}, R.~V., \& {Fabian}, A.~C. 2007, \mnras, 381, 1235,
  \dodoi{10.1111/j.1365-2966.2007.12328.x}

\bibitem[{{Vasudevan} {et~al.}(2014){Vasudevan}, {Mushotzky}, {Reynolds},
  {Fabian}, {Lohfink}, {Zoghbi}, {Gallo}, \& {Walton}}]{vas14}
{Vasudevan}, R.~V., {Mushotzky}, R.~F., {Reynolds}, C.~S., {et~al.} 2014, \apj,
  785, 30, \dodoi{10.1088/0004-637X/785/1/30}

\bibitem[{{Vaughan} {et~al.}(2003){Vaughan}, {Edelson}, {Warwick}, \&
  {Uttley}}]{vau03}
{Vaughan}, S., {Edelson}, R., {Warwick}, R.~S., \& {Uttley}, P. 2003, \mnras,
  345, 1271, \dodoi{10.1046/j.1365-2966.2003.07042.x}

\bibitem[{{Verner} {et~al.}(1996){Verner}, {Ferland}, {Korista}, \&
  {Yakovlev}}]{ver96}
{Verner}, D.~A., {Ferland}, G.~J., {Korista}, K.~T., \& {Yakovlev}, D.~G. 1996,
  \apj, 465, 487, \dodoi{10.1086/177435}

\bibitem[{{Walter} \& {Fink}(1993)}]{wal93}
{Walter}, R., \& {Fink}, H.~H. 1993, \aap, 274, 105

\bibitem[{{Walton} {et~al.}(2013){Walton}, {Nardini}, {Fabian}, {Gallo}, \&
  {Reis}}]{wal13}
{Walton}, D.~J., {Nardini}, E., {Fabian}, A.~C., {Gallo}, L.~C., \& {Reis},
  R.~C. 2013, \mnras, 428, 2901, \dodoi{10.1093/mnras/sts227}

\bibitem[{{Walton} {et~al.}(2014){Walton}, {Risaliti}, {Harrison}, {Fabian},
  {Miller}, {Arevalo}, {Ballantyne}, {Boggs}, {Brenneman}, {Christensen},
  {Craig}, {Elvis}, {Fuerst}, {Gandhi}, {Grefenstette}, {Hailey}, {Kara},
  {Luo}, {Madsen}, {Marinucci}, {Matt}, {Parker}, {Reynolds}, {Rivers}, {Ross},
  {Stern}, \& {Zhang}}]{wal14}
{Walton}, D.~J., {Risaliti}, G., {Harrison}, F.~A., {et~al.} 2014, \apj, 788,
  76, \dodoi{10.1088/0004-637X/788/1/76}

\bibitem[{{Walton} {et~al.}(2016){Walton}, {Tomsick}, {Madsen}, {Grinberg},
  {Barret}, {Boggs}, {Christensen}, {Clavel}, {Craig}, {Fabian}, {Fuerst},
  {Hailey}, {Harrison}, {Miller}, {Parker}, {Rahoui}, {Stern}, {Tao}, {Wilms},
  \& {Zhang}}]{wal16}
{Walton}, D.~J., {Tomsick}, J.~A., {Madsen}, K.~K., {et~al.} 2016, \apj, 826,
  87, \dodoi{10.3847/0004-637X/826/1/87}

\bibitem[{{Wilms} {et~al.}(2000){Wilms}, {Allen}, \& {McCray}}]{wil00}
{Wilms}, J., {Allen}, A., \& {McCray}, R. 2000, \apj, 542, 914,
  \dodoi{10.1086/317016}

\bibitem[{{Woo} \& {Urry}(2002)}]{woo02}
{Woo}, J.-H., \& {Urry}, C.~M. 2002, \apj, 579, 530, \dodoi{10.1086/342878}

\bibitem[{{Xu} {et~al.}(2017){Xu}, {Balokovi{\'c}}, {Walton}, {Harrison},
  {Garc{\'{\i}}a}, \& {Koss}}]{xu17}
{Xu}, Y., {Balokovi{\'c}}, M., {Walton}, D.~J., {et~al.} 2017, \apj, 837, 21,
  \dodoi{10.3847/1538-4357/aa5df4}

\bibitem[{{Yaqoob} {et~al.}(2007){Yaqoob}, {Murphy}, {Griffiths}, {Haba},
  {Inoue}, {Itoh}, {Kelley}, {Kokubun}, {Markowitz}, {Mushotzky}, {Okajima},
  {Ptak}, {Reeves}, {Serlemitsos}, {Takahashi}, \& {Terashima}}]{yaq07}
{Yaqoob}, T., {Murphy}, K.~D., {Griffiths}, R.~E., {et~al.} 2007, \pasj, 59,
  283

\bibitem[{{Zdziarski} {et~al.}(1996){Zdziarski}, {Johnson}, \&
  {Magdziarz}}]{zdz96}
{Zdziarski}, A.~A., {Johnson}, W.~N., \& {Magdziarz}, P. 1996, \mnras, 283,
  193, \dodoi{10.1093/mnras/283.1.193}

\bibitem[{{{\.Z}ycki} {et~al.}(1999){{\.Z}ycki}, {Done}, \& {Smith}}]{zyc99}
{{\.Z}ycki}, P.~T., {Done}, C., \& {Smith}, D.~A. 1999, \mnras, 309, 561,
  \dodoi{10.1046/j.1365-8711.1999.02885.x}

\end{thebibliography}
%
%
%
%
\end{document}